\documentclass[manuscript]{aastex}
\usepackage{color}
\usepackage{amsmath,graphics,graphicx}
\usepackage{rotating}
\usepackage{lscape}

\newcommand{\lapprox} {\, \lower3pt\hbox{$\sim$}\llap{\raise2pt\hbox{$<$}}\,}
\newcommand{\gapprox} {\, \lower3pt\hbox{$\sim$}\llap{\raise2pt\hbox{$>$}}\,}

\shorttitle{Collisional Relaxation in Warm Plasma}
\shortauthors{Kontar et al.}

\begin{document}

\title{COLLISIONAL RELAXATION OF ELECTRONS IN A WARM PLASMA AND ACCELERATED NONTHERMAL ELECTRON SPECTRA IN SOLAR FLARES}

\author{Eduard P. Kontar\altaffilmark{1}, Natasha L. S. Jeffrey\altaffilmark{1}, A. Gordon Emslie\altaffilmark{2} and N. H. Bian\altaffilmark{1}}

\altaffiltext{1}{School of Physics \& Astronomy, University of Glasgow, G12 8QQ, Glasgow, Scotland, United Kingdom \\ email: eduard.kontar@astro.gla.ac.uk}

\altaffiltext{2}{Department of Physics \& Astronomy, Western Kentucky University, Bowling Green, KY 42101}

\begin{abstract}
Extending previous studies of nonthermal electron transport in solar flares which include the effects of collisional energy diffusion and thermalization of fast electrons, we present an analytic method to infer more accurate estimates of the accelerated electron spectrum in solar flares from observations of the hard X-ray spectrum. Unlike for the standard cold-target model, the spatial characteristics of the flaring region, especially the necessity to consider a finite volume of hot plasma in the source, need to be taken into account in order to correctly obtain the injected electron spectrum from the source-integrated electron flux spectrum (a quantity straightforwardly obtained from hard X-ray observations). We show that the effect of electron thermalization can be significant enough to nullify the need to introduce an {\it ad hoc} low-energy cutoff to the injected electron spectrum in order to keep the injected power in non-thermal electrons at a reasonable value. Rather the suppression of the inferred low-energy end of the injected spectrum compared to that deduced from a cold-target analysis allows the inference from hard X-ray observations of a more realistic energy in injected non-thermal electrons in solar flares.
\end{abstract}


\section{Introduction}\label{intro}

The copious hard X-ray emission produced during solar flares is the main evidence for the acceleration of a large number of suprathermal electrons in such events \citep[see, e.g.,][]{2011SSRv..159..107H,2011SSRv..159..301K}. These accelerated electrons propagate through the surrounding plasma, where electron-ion collisions give rise to the observed bremsstrahlung hard X-rays. The collisions with ambient particles, principally other electrons \citep[e.g.,][]{1972SoPh...26..441B,1978ApJ...224..241E}, are responsible for energy transfer from the accelerated particles to the ambient plasma.

In a cold-target approximation, where the dynamics of hard X-ray emitting electrons are dominated by systematic energy loss \citep[e.g.,][]{1971SoPh...18..489B}, only a small fraction ($\sim 10^{-5}$) of the energy in the accelerated electrons is emitted as bremsstrahlung hard X-rays. An observed hard X-ray flux thus translates into a much higher accelerated electron energy content, to the extent that the energy in such accelerated electrons can be comparable to the energy in the stressed pre-flare magnetic field and to the total energy radiated during the flare \citep[see][]{2004JGRA..10910104E,2005JGRA..11011103E,2012ApJ...759...71E}.

Solar flare hard X-ray spectra $I(\epsilon)$ (photons~cm$^{-2}$~s$^{-1}$~keV$^{-1}$ at the Earth) in the nonthermal domain are typically quite steep, with power-law forms $I(\epsilon) \sim \epsilon^{-\gamma}$ and spectral indices $\gamma \simeq (3 - 6)$ \citep[see, e.g.,][for a review]{2011SSRv..159..301K}.  Using a cold target model (that retains only the effect of energy loss in the dynamics of emitting electrons) requires that the injected electron flux spectra $F_0(E_0)\propto E_0^{-\delta}$ (electrons~cm$^{-2}$~s$^{-1}$~keV$^{-1}$) are similarly steep, with a power-law indices $\delta = \gamma+1$ \citep[e.g.,][]{1971SoPh...18..489B}. Since the total injected energy flux ${\cal E} = \int_0^\infty E_0 \, F_0(E_0) \, dE_0$ diverges at the lower limit for such steep power laws, the concept of a ``low-energy cutoff'' $E_{\rm c}$ is frequently assumed in order to keep the value of ${\cal E} = \int_{E_{\rm c}}^\infty E_0 \, F_0(E_0) \, dE_0$ finite.  Indeed the value of $E_{\rm c}$ is usually \citep{2003ApJ...595L..97H} taken to be the maximum value consistent with the hard X-ray data, resulting in the minimum energy in the nonthermal electrons that is consistent\footnote{Because of the dominance of the (even steeper) thermal hard X-ray component at low photon energies $\epsilon$, lower values of $E_{\rm c}$ are permitted by the data but result in unjustifibaly large values of the injected energy flux ${\cal E} = \int_{E_{\rm c}}^\infty E_0 \, F_0(E_0) \, dE_0$.} with the data \citep[see, e.g.,][]{2003ApJ...595L..97H,2011SSRv..159..107H}.

Observations indicate that fast electrons are accelerated in, and subsequently move through, the hot ($\sim$$10^7$~K) flaring plasma in the corona. In some cases, the coronal density is sufficiently high to arrest the electrons wholly within this coronal region, resulting in a thick-target coronal source \citep[e.g.,][]{2004ApJ...603L.117V,2008ApJ...673..576X}. In other cases, the coronal density is sufficiently low that the electrons emerge, with somewhat reduced energy, from the coronal region and then impact the relatively cool ($\sim$$10^4$~K) gas in the chromosphere, producing the commonly observed ``footpoint-dominated'' flares.  The plasma in the corona and in the preflare chromosphere is rapidly heated to temperatures in excess of $\sim$$10^7$~K, producing the commonly-observed soft X-ray emission over the extent of the flaring loop. This heating (or direct heating in magnetic energy release) necessarily renders a significant portion of the target ``warm''; i.e., such that, for an appreciable portion of the injected electron population, the injected energy $E_0$ is comparable to the thermal energy $kT$, where $k$ is Boltzmann's constant and $T$ the target temperature.

\citet{2003ApJ...595L.119E} included consideration of the finite temperature of the target in modifying the systematic energy loss rate of the accelerated electrons; such considerations come into play as the electron energies $E$ approach a few $kT$.  He showed that this resulted in a lowering of the required energy flux ${\cal E}$ of injected electrons for a prescribed hard X-ray intensity.  Further, \citet{2005A&A...438.1107G} have emphasized the role of energy diffusion, which is also important at energies of a few $kT$ and is a necessary ingredient for describing thermalization of the fast electrons in a warm target. \citet{2014ApJ...787...86J} showed that the transport of electrons is more complicated than assumed by either \citet{2003ApJ...595L.119E} or \citet{2005A&A...438.1107G}, inasmuch as the effects of diffusion in both energy and space must be included in a self-consistent analysis of electron transport in a warm target.

In this paper we follow \citet{2005A&A...438.1107G}, \citet{2010PhPl...17k2313G}, and \citet{2014ApJ...787...86J}. We show that thermalization of fast electrons in a warm ambient target significantly changes the evolution of the electron energy spectrum compared to that in a finite temperature target analysis that neglects diffusion in energy \citep{2003ApJ...595L.119E}, and even more so compared to the standard cold target model \citep[e.g.,][]{1971SoPh...18..489B}. We also emphasize that the appropriate treatment of the thermalization of electrons injected into a warm target results into a higher hard X-ray yield per electron, so that significantly fewer supra-thermal accelerated electrons are required to be injected into the target in order to produce a given observed hard X-ray flux. Moreover, our analysis shows that, contrary to the case of a cold target (in which the spatially-integrated hard X-ray yield is independent of the density profile of the target), in a warm target one \textit{does} need to take into account the spatial characteristics of the emitting region, in particular the extent of the warm target compared to that of the overall flaring region.

The outline of the paper is as follows. First, we quantitatively evaluate the effects of thermalization on electron transport, using both analytical (Sections~\ref{analysis-analytical-forward} and~\ref{analysis-analytical}) and numerical (Section~\ref{sim_method}) methods, which we compare in Section~\ref{comparison}. We find that warm-target thermalization of electrons considerably reduces the flux of injected electrons compared to that obtained in a cold-target model that assumes a low-energy cutoff below the thermalization limit derived here, and even compared to that in a model \citep{2003ApJ...595L.119E} that includes warm-target effects on the secular energy loss rate but neglects diffusion in energy. In Section~\ref{discussion} we summarize the results and point out that the magnitude of the effect is sufficiently large so that a low-energy cutoff $E_{\rm c}$ in the injected power ${\cal P} = A \, \int E_0 \, F_0(E_0) \, dE_0$ (where $A$ (cm$^2$) is the injection area) is a natural consequence, rather than an {\it ad hoc} assumption adopted {\it a posteriori}. We therefore urge discontinuance of cold target modeling for a warm plasma and its resultant need to impose an {\it ad hoc} low-energy cutoff. Rather we encourage the computation of the injected electron spectrum through a more realistic model which includes both the effects of friction and diffusion in the evolution of the electron distribution as a whole, letting the low-energy end of the injected spectrum (and hence the injected power in nonthermal electrons) be determined naturally from the underlying physics.

\section{RELATION BETWEEN THE SOURCE-INTEGRATED SPECTRUM AND THE INJECTED ELECTRON SPECTRUM}\label{relation}

\citet{2003ApJ...595L.115B} have shown that the bremsstrahlung hard X-ray spectrum $I(\epsilon)$ (photons~cm$^{-2}$~s$^{-1}$~keV$^{-1}$ at the Earth) is related to the \textit{source-integrated electron flux spectrum} $\langle nVF \rangle (E)$ (electrons~cm$^{-2}$~s$^{-1}$~keV$^{-1}$) by

\begin{equation}\label{xray-electron}
I(\epsilon) = \frac{1}{4 \pi R^2} \int_\epsilon^\infty \int_V n({\bf r}) \, F(E,{\bf r}) \, dV \, \sigma(\epsilon, E) \, dE
= \frac{1}{4 \pi R^2} \int_\epsilon^\infty \langle nVF \rangle (E) \, \sigma(\epsilon, E) \, dE \,\,\, .
\end{equation}
Here $n$ is the ambient proton density (cm$^{-3}$) at position ${\bf r}$, $V$ (cm$^3$) is the source volume, $R=1$~AU, and $\sigma(\epsilon,E)$ (cm$^2$~keV$^{-1}$) is the bremsstrahlung cross-section, differential in photon energy $\epsilon$. For a stratified one-dimensional target,

\begin{equation}\label{fbar-def-1d}
\langle nVF \rangle \, (E) = A \, \int_V F(E,N) \, dN \,\,\, ,
\end{equation}
where $F(E,N)$ is the electron flux spectrum, $A$ is the cross-sectional area in the direction of electron propagation ${\hat {\bf z}}$, and $N = \int n(z) \, dz$ is the column density (cm$^{-2}$).

For a specified form of the bremsstrahlung cross-section $\sigma(\epsilon, E)$, Equation~(\ref{xray-electron}) uniquely determines $\langle nVF \rangle (E)$ from observations of $I(\epsilon)$; no assumptions are required regarding the dynamics of the emitting electrons. On the other hand, relating $\langle nVF \rangle (E)$ to the injected electron flux spectrum $F_0(E_0)$ (electrons~cm$^{-2}$~s$^{-1}$~keV$^{-1}$) \textit{does} require assumptions on the dynamics of electrons in the target, and any change in the energy loss (or gain) compared to the cold-target value \citep{1972SoPh...26..441B,1978ApJ...224..241E} will change this relation \citep[e.g.,][]{2009A&A...508..993B, 2012A&A...539A..43K}.

\subsection{Form of the injected spectrum for a prescribed source-integrated electron spectrum}\label{analysis-analytical-forward}

\citet{2005A&A...438.1107G} have found that in a warm target \textit{both} friction and diffusion due to Coulomb collisions affect the evolution of the ensemble of accelerated particles. Therefore they are both important in establishing the relationship between the source-integrated electron spectrum $\langle nVF \rangle (E)$ and the injected electron spectrum $F_0(E_{0})$.  To describe such an environment, \citet{2014ApJ...787...86J} used the Fokker-Planck equation
\begin{eqnarray}\label{eq:fp}
\mu \, \frac{\partial F}{\partial z} &= & 2 K n \left \{ \frac{\partial}{\partial E} \left [ G \left (\sqrt{\frac{E}{k_B T}} \, \right ) \, \frac{\partial F}{\partial E} +\frac{1}{E} \, \left ( \frac{E}{k_B T} - 1 \right ) \, G \left (\sqrt{\frac{E}{k_B T}} \, \right ) \, F \right ] \right .
+ \cr
& + & \left . \frac{1}{8E^2} \, \frac{\partial}{\partial \mu} \left [ (1-\mu^{2}) \, \left ( {\rm erf} \left ( \sqrt{\frac{E}{k_B T}} \, \right ) -G\left (\sqrt{\frac{E}{k_B T}} \, \right ) \right ) \frac{\partial F}{\partial \mu} \right ]\right \} \, + \, F_0(E) \, \delta(z) \,\,\, .
\end{eqnarray}
Here $F(E,\mu,z)$ is the local electron flux at position $z$ along the guiding magnetic field, energy $E$, and pitch-angle cosine $\mu$. $F_0(E) \, \delta(z)$ is the source of accelerated
electrons which are injected into the target at $z=0$, and $K = 2 \pi e^4 \Lambda$ is the collision parameter, with $\Lambda$ being the Coulomb logarithm. $G(u)$ is the Chandrasekhar function, given by

\begin{equation}\label{eq:gchatext}
G(u)=\frac{{\rm erf}(u)-u \, {\rm erf}^{'}(u)}{2u^{2}}  \,\,\, ,
\end{equation}
where ${\rm erf}(u)\equiv (2/\sqrt{\pi})\int\limits_{0}^{u}\exp(-t^2) \, dt$ is the error function.

Averaging Equation~(\ref{eq:fp}) over pitch-angle and integrating over the emitting volume $V = \int A \, dz$ \citep[as in][]{2014ApJ...780..176K} gives the relationship between the source-integrated electron spectrum $\langle nVF \rangle (E)$ and the (pitch-angle averaged) injected electron spectrum $F_0(E_{0})$, viz.

\begin{equation}\label{eq:nvF_full}
F_0(E_{0}) = - \frac{2K} { A} \, \frac{d}{dE} \left [ G \left (\sqrt{\frac{E}{kT}} \, \right ) \, \left \{ \frac{d \langle nVF \rangle (E)}{dE}
+\frac{1}{E} \, \left ( \frac{E}{kT} - 1 \right ) \, \langle nVF \rangle (E) \right \} \right ]_{E=E_0} \,\,\, .
\end{equation}
It should be noted that substitution of a Maxwellian at temperature $T$: $\langle nVF \rangle (E) \sim E \, \exp(-E/kT)$ results in the right side of Equation~(\ref{eq:nvF_full}) vanishing identically, so that the required injected flux (source function) $F_0(E_{0})$ corresponding to a Maxwellian form of $\langle nVF \rangle (E)$ is zero, as it should be in a dynamical model which is solely governed by collisional effects.  It also follows that adding a Maxwellian of arbitrary size to an inferred $\langle nVF \rangle (E)$ has no effect on the resulting $F_0(E_0)$.

Let us now compare the constitutive relation (\ref{eq:nvF_full}) between $F_{0}(E_{0})$ and $\langle nVF \rangle (E)$ with those used by previous authors. Neglecting the second-order energy diffusion term in Equation~(\ref{eq:nvF_full}) gives

\begin{equation}\label{f0-fbar}
F_0(E_0) = -\frac{2K}{A} \, \frac{d}{dE} \left [ \frac{1}{E} \, \left ( \frac{E}{kT} - 1 \right ) \, G \left (\sqrt{\frac{E}{kT}} \, \right ) \, \langle nVF \rangle (E) \right ]_{E=E_0} \,\,\, .
\end{equation}
This is the relationship in the finite-temperature diffusionless ``warm-target'' model proposed by \citet{2003ApJ...595L.119E}.

As a final simplification, we may assume that the target is ``cold,'' i.e., $E \gg kT$ \citep{1971SoPh...18..489B}. In this regime, $G(u) \rightarrow 1/2u^2$ (Equation~(\ref{eq:gchatext})), resulting in the compact expression

\begin{equation}\label{fbar-f0-cold}
F_0(E_0) = - \frac{K}{A} \, \frac{d} {dE} \left [ \frac{\langle nVF \rangle (E)}{ E} \right ]_{E=E_0}
\end{equation}
previously used by \citet{1984ApJ...279..882E} and \citet{1988ApJ...331..554B}.

Although the relationship~(\ref{fbar-f0-cold}) has been used by a number of authors to deduce the injected flux spectrum $F_0(E_{0})$ from the source-integrated electron spectrum $\langle nVF \rangle (E)$, we stress that the form~(\ref{eq:nvF_full}) is a much more accurate representation of the true relationship in a warm target. While the relationship~(\ref{f0-fbar}), first suggested by \citet{2003ApJ...595L.119E}, does take into account the effect of finite target temperature on the energy loss rate, it neglects the concomitant energy diffusion responsible for thermalization in a warm target, which, as we shall see, is a finite temperature effect of even greater importance.

\begin{figure*}[pht]
\centering
\includegraphics[width=0.6\linewidth]{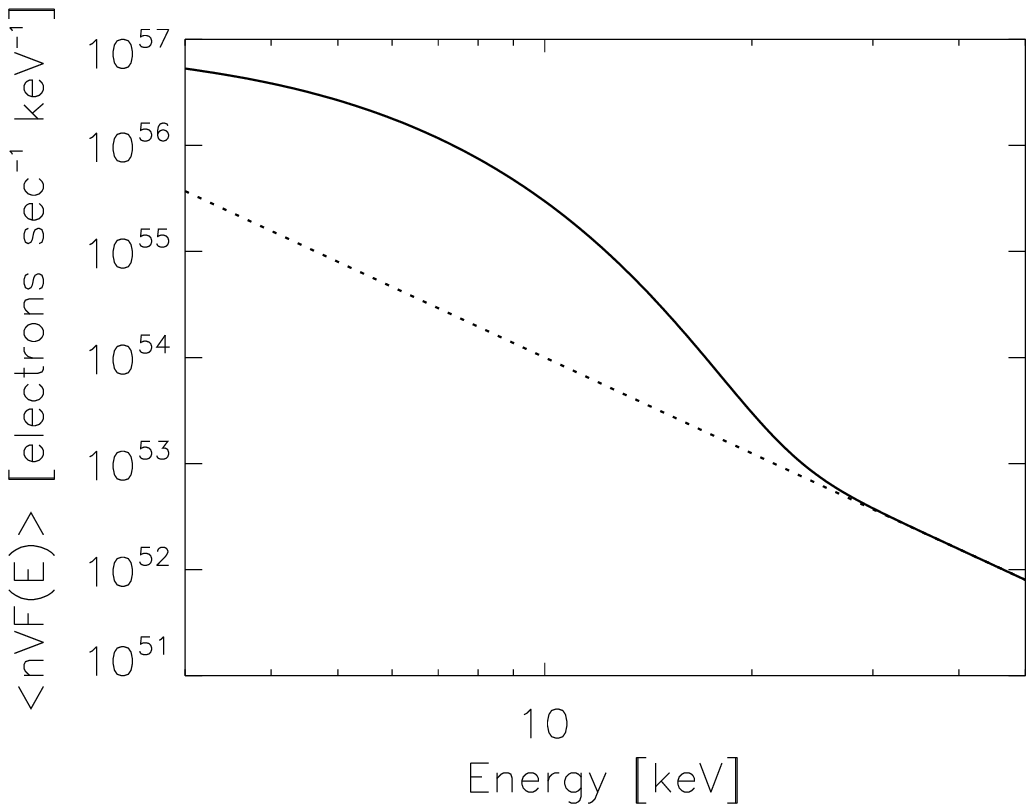}\\
\includegraphics[width=0.6\linewidth]{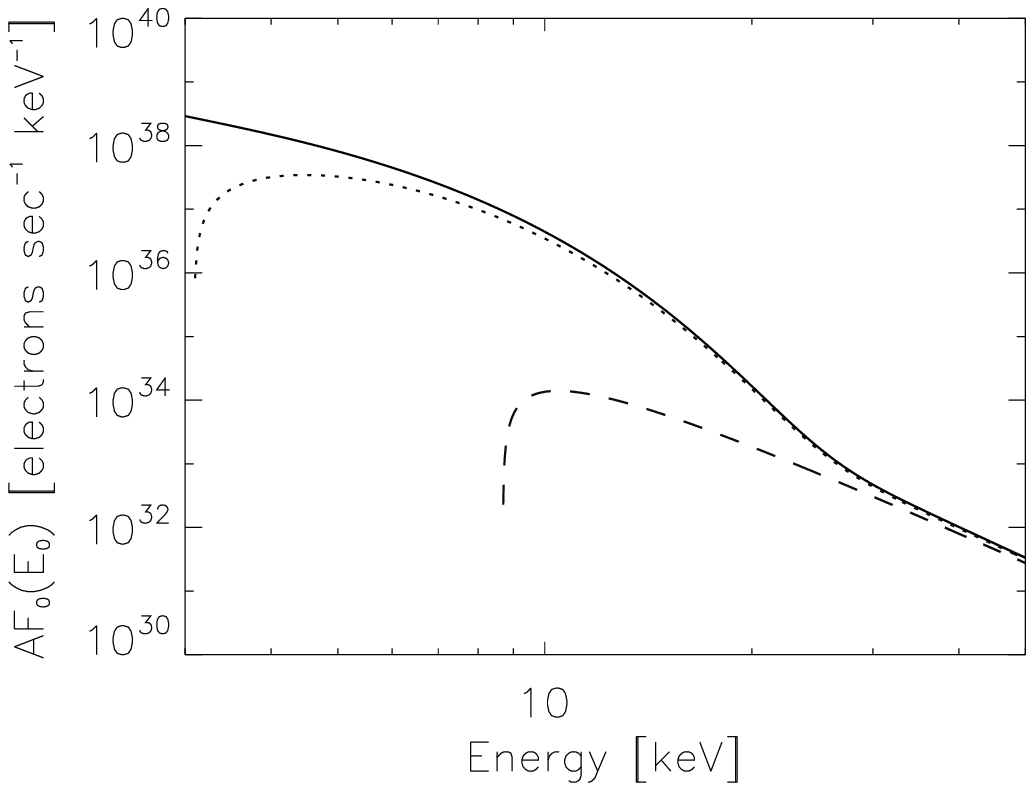}
\caption{Top: Mean electron flux prescribed by $\langle nVF \rangle (E)=10^{48} \,F_{M}(E)+10^{54}(E/10~{\rm keV})^{-3}$ (electrons~s$^{-1}$~keV$^{-1}$), where $F_M(E)$ is the Maxwellian flux spectrum normalized to unity with $kT = 1.75$~keV. Bottom: Forms of the injected spectrum $A \, F_0(E_0)$ recovered for a prescribed top-panel $\langle nVF \rangle (E)$ using different model assumptions: \textit{solid line} - cold target (Equation~(\ref{fbar-f0-cold})), \textit{dotted-line}: non-diffusional warm target (Equation~(\ref{f0-fbar})), and \textit{dashed line}: diffusional warm target (Equation~(\ref{eq:nvF_full})).}
\label{fig:different_f0}
\end{figure*}

In order to evaluate the impact of energy diffusion on the relationship between the injected flux spectrum $F_0(E_0)$ and the source-integrated electron spectrum $\langle nVF \rangle (E)$, we performed the following simple analysis. First, we assumed a hard X-ray spectrum consisting of combination of a Maxwellian with emission measure EM = $1 \times 10^{48}$~cm$^{-3}$ and temperature $T = 20$~MK ($kT \simeq 1.75$~keV) and a nonthermal spectrum with power-law form $\langle nVF \rangle (E) = 10^{54} (E/10~{\rm keV})^{-\zeta}; E \ge E_*$, corresponding to a moderately large solar flare.  We selected $\zeta =3$ and $E_* = 3$~keV (see Figure \ref{fig:different_f0}); as we shall show, the results are insensitive to the value of $E_*$ as long as it is significantly less than about 10~keV.  Then, using a model warm target with $kT=1.75$~keV, we used Equation~(\ref{eq:nvF_full}) to determine the corresponding form of the areally-integrated injected spectrum $A \, F_0(E_0)$ (electrons~s$^{-1}$~keV$^{-1}$). We then compared this with the forms of $A \, F_0(E_0)$ resulting from application of Equations~(\ref{f0-fbar}) (non-diffusional warm target) and~(\ref{fbar-f0-cold}) (cold target), respectively.

Figure~\ref{fig:different_f0} shows the results, with the salient features summarized below.

\begin{enumerate}

\item At high energies, the cold target result (solid line) is, as may be verified analytically from Equation~(\ref{fbar-f0-cold}), a power law with index $\delta = \zeta+2 = 5$.  Extending this spectrum down to the $3$~keV lower boundary of the $\langle nVF \rangle (E)$ spectrum requires an injection rate ${\dot N} = A \, \int_{3 \, {\rm keV}}^\infty F_0(E_0) \, dE_0 \simeq 10^{37}$ electrons~s$^{-1}$ and a corresponding power ${\cal P} = A \, \int_{3 \, {\rm keV}}^\infty E_0 \, F_0(E_0) \, dE_0 \simeq 7\times 10^{28}$~erg~s$^{-1}$. Because of this steep spectral form, imposition of an arbitrary low-energy cutoff $E_{\rm c}$ is necessary to keep the injected power ${\cal P} = A \int_{E_{\rm c}}^\infty E_0 \, F_0(E_0) \, dE_0$ at an acceptably small value.

\item The result for a diffusionless warm target (Equation~(\ref{f0-fbar})) does flatten off below a few keV and imposes a lower bound to $E_{\rm c}$ at $E_0 \simeq kT$ \citep[][bottom panel of Figure \ref{fig:different_f0}]{2003ApJ...595L.119E}.

\item Application of the diffusional warm target model (Equation \ref{eq:nvF_full}) causes the Maxwellian component in the hard X-ray spectrum to have a corresponding null injected spectrum $F_0(E_0)$; only the nonthermal power-law component is associated with a finite $F_0(E_0)$.  Further, the inferred $F_0(E_0)$ has a rather abrupt cutoff at $E_0 = E_{\rm c,eff} \simeq 9$~keV, and the $F_0(E_0)$ is also noticeably flatter than the cold-target result up to energies as high as twice this cutoff value.  Because of these features, this model requires only ${\dot N} = A \int_0^\infty F_0(E_0) \, dE_0 \simeq 2\times 10^{35}$ electrons~s$^{-1}$,  a factor of $\sim$50 lower than in a cold target model above 3~keV. Similarly, the required power ${\cal P} = A \int_0^\infty E_0 \, F_0(E_0) \, dE_0 \simeq 4 \times 10^{27}$~erg~s$^{-1}$ is a factor of $\sim$20 lower than the power above the {\it a priori} imposed cutoff energy of $3$~keV. We note that the form of $F_0(E_0)$ allows the integrals for the injected rate and power to be taken over the entire $E_0$ range $[0, \infty)$.

\end{enumerate}

It must be stressed that the effective cutoff in the form of $F_0 (E_0)$ at $E_0 = E_{\rm c,eff} \simeq 9$~keV results naturally from the physics, rather than being imposed \textit{a posteriori} in order to keep the injected power at a minimum level \citep[e.g. \textit{maximum} value of the cut-off, ][]{2003ApJ...595L..97H}.  To see this, we assume a source-integrated mean electron flux\footnote{Note that adding a Maxwellian with temperature $kT$ and arbitrary magnitude does not affect the result. So one can assume that we have a Maxwellian plus power-law mean flux spectrum, as is often observed in solar flares.} $\langle nVF \rangle (E) \propto E^{-\zeta}$, and set the right hand side of Equation~(\ref{eq:nvF_full}) to zero:

\begin{equation}\label{eq:nvF_full_der1}
\frac{d}{dE} \left [ G \left (\sqrt{\frac{E}{kT}} \, \right ) \, \left \{ \frac{ d E^{-\zeta}}{dE}
+\frac{1}{E} \, \left ( \frac{E}{kT} - 1 \right ) \, E^{-\zeta} \right \} \right ]_{E=E_0} = 0 \,\,\, .
\end{equation}
For $E\gg kT$, $G \left (\sqrt{{E}/{kT}}\right )\simeq kT/2E$ and we can simplify further:

\begin{equation}\label{eq:nvF_full_der2}
 \frac{d}{dE} \left[ E^{-\zeta-2}   \,\left\{\frac{E}{kT} - (\zeta + 1) \, \right\} \right ]_{E=E_0} = 0 \,\,\, ,
\end{equation}
which vanishes at an effective cutoff energy $E_0 = E_{\rm c,eff}$, where

\begin{equation}\label{eq:E_cut_analytic}
E_{\rm c,eff} = (\zeta+2) kT \,\,\, .
\end{equation}
For $kT=1.75$~keV and $\zeta =3$, Equation~(\ref{eq:E_cut_analytic}) gives $E_{\rm c,eff} \simeq 9$~keV, in agreement with the results shown in Figure~\ref{fig:different_f0}.  The effective low energy cut-off $E_{\rm c,eff}$ could be as large as $24$~keV for a relatively soft spectrum with $\zeta =6$ and a relatively hot ($kT \simeq 3$~keV) flaring plasma; such a value of $E_{\rm c,eff}$ is comparable to the typical (upper bound) values used in the literature \citep[e.g.,][]{2011SSRv..159..107H}.

We further note that below the effective cutoff energy $E_{\rm c, eff}$, the values of $F_0(E_0)$ inferred from Equation~(\ref{eq:nvF_full}) are {\it negative}. This unphysical situation is forced by the assumption of a strict power-law form of $\langle n VF \rangle (E)$ down to low energies $E_*$. In a real flare (with positive semidefinite $F_0(E_0)$), the form of $\langle n VF \rangle (E)$ will necessarily deviate from this power-law form and instead transition to a Maxwellian form at low energies.  This will be studied in the following subsection.

\subsection{Form of the source-integrated electron spectrum for a prescribed injected spectrum}\label{analysis-analytical}

Considerable insight into the effects of the energy diffusion term can be obtained by carrying out the reverse procedure to the previous section, i.e., solving Equation~(\ref{eq:nvF_full}) for $\langle nVF \rangle (E)$ for a given injected spectrum $F_0(E_0)$, and comparing the results with numerical simulations based on the full Fokker-Planck equation~(\ref{eq:fp}).

A first integral of Equation~(\ref{eq:nvF_full}) is

\begin{equation}\label{eq: first-integral}
\frac{d \langle nVF \rangle (E)}{dE} +  \left ( \frac{1}{kT} - \frac{1}{E} \right ) \, \langle nVF \rangle (E)
= \frac{A}{ 2K G \left ( \sqrt{\frac{E}{kT}}\right)} \, \int_E^\infty F_0(E_0) \, dE_0 \,\,\, .
\end{equation}
Adding an integrating factor $e^{E/kT}$ allows this to be written

\begin{equation}\label{eq: second-integral}
\frac{d}{dE} \left ( \frac{e^{E/kT}}{ E} \, \langle nVF \rangle (E) \right ) = \left ( \frac{1}{2K} \right ) \,
\frac{e^{E/kT}}{E \, G \left ( \sqrt{\frac{E}{kT}} \, \right ) } \, \int_E^\infty A \, F_0(E_0) \, dE_0 \,\,\, ,
\end{equation}
which can in turn be straightforwardly integrated from a lower limit $E_{\rm min}$ (see next subsection) to $E$ to give the desired expression:

\begin{equation}\label{eq: fbar-solution}
\langle nVF \rangle (E) = \frac{1}{2K} \, E \, e^{-E/kT} \, \int_{E_{\rm min}}^{E}
\frac{e^{E^\prime/kT} \, dE^\prime }{E^\prime \, G \left ( \sqrt{\frac{E^\prime}{kT}} \, \right ) } \,
\int_{E^\prime}^\infty A \, F_0(E_0) \, dE_0 \,\,\, .
\end{equation}

It is important to notice that for $E_{\rm min} = 0$ the solution (\ref{eq: fbar-solution}) is \textit{infinite}; this can be seen by considering the limit for $E' \rightarrow 0$ and noting that $G(u)\rightarrow u$ as $u\rightarrow 0$, so that the outer integrand $\sim {E^\prime}^{-3/2}$ and hence the integral diverges as $E_{\rm min}\rightarrow 0$. Physically, this divergence arises from the fact that a finite stationary solution does not exist because we have a constant source of particles, no sink of particles, and a Fokker-Planck collisional operator that conserves the total number of particles. Hence, in the absence of escape, the number of electrons in the stationary state grows indefinitely. However, as we shall now show, the finite extent of the warm plasma region leads to an escape term which results in a non-zero lower limit $E_{\rm min}$ and hence a finite number of electrons.

\subsubsection{Effects of escape from a finite-extent hot plasma region}\label{finite_sol}

In a standard flare scenario, the electrons accelerated in the hot corona propagate downwards towards the dense chromospheric layers. Recent high resolution observations \citep[e.g.,][]{2011A&A...533L...2B} confirm this picture and show that the $\sim$40~keV electrons form HXR footpoints (height $\sim$1~Mm above the photosphere), below the position of the EUV emission (height $\sim$(2-3)~Mm) produced as a result of energy deposition by lower energy electrons or thermal conduction.  From a theoretical standpoint, as shown by a number of authors \citep[e.g.,][]{1984ApJ...279..896N,1989ApJ...341.1067M}, the injection of electrons into the top layers of the chromosphere could result in a rapid rise of temperature of the preflare chromospheric material from $T \simeq 10^4$~K ($kT \simeq 0.001$~keV) through the region of radiative instability \citep{1969ApJ...157.1157C}, to a temperature\footnote{The flare coronal temperature is effectively limited to a few $\times 10^7$~K because of the very strong \citep[$\propto T^{7/2}$;][]{1962pfig.book.....S} dependence of thermal conductive losses on temperature.} $T \simeq (1 - 3) \times 10^7$~K ($kT \simeq (1-3)$~keV) in a few seconds or slower depending on the power injected by the beam. However, below a certain height, the density of the chromosphere is so high that the plasma can effectively radiate  away the injected electron power. Therefore, both observations and theoretical expectations show that any realistic flare volume will be composed of two main parts: a portion of the target (low chromosphere) that is cold ($kT \simeq 0.001$~keV $\ll E$) and another portion that is ``warm'' ($E \sim kT$).  In the latter volume the effect of diffusional thermalization of the fast electrons is important.

Therefore let us consider the warm target portion of the loop, with density $n$ and finite length $L$ and considered sufficiently ``thick'' to thermalize the low-energy region of the injected suprathermal particles, while higher-energy electrons may still propagate through this warm target without significant energy loss or scattering.  The spatial transport of such electrons in the warm target is dominated by collisional diffusion \citep[similarly to Equation (22) in][]{2014ApJ...780..176K}, giving

\begin{equation}\label{eq:fpe}
\frac{1}{v}\frac{\partial }{\partial z} \left ( D_c\frac{\partial F}{\partial z} \right ) = 2 K n \, \frac{\partial}{\partial E} \left [ G \left (\sqrt{\frac{E}{kT}} \, \right ) \, \frac{\partial F}{\partial E} +\frac{1}{E} \, \left ( \frac{E}{kT} - 1 \right ) \, G \left (\sqrt{\frac{E}{kT}} \, \right ) \, F \right ] + \, F_0(E) \, \delta(z) \,\,\, ,
\end{equation}
where $D_c$ is the collisional diffusion coefficient. Since the collisional operator, the second term in Equation~(\ref{eq:fpe}), conserves the number of electrons, the effect of injection of electrons into such a coronal region at a rate ${\dot N}$ ($s^{-1}$) is to increase $N$, the number of electrons in the target.  In a stationary state the number of electrons in the target is balanced between injection and diffusive escape of thermalized electrons with energy $E\simeq kT$ out of the warm part of the target, at a rate

\begin{equation}\label{esc-diffusion}
\dot{N} = \frac{\partial}{\partial z} \, \left ( D_c \, \frac{\partial N}{\partial z} \right ) \simeq \frac{2 D_c}{L^2} \, N = \frac{2 kT \, \tau_e}{L^2 \, m_e} \, N \,\,\, ,
\end{equation}
where

\begin{equation}\label{tau-e}
\tau_e = 3 \, \sqrt{\frac{\pi m_e}{8}} \, \frac{(kT)^{3/2}}{Kn}
\end{equation}
is the electron collision time \citep[e.g.][]{1981phki.book.....L,1983nrl..reptQ....B}. This balance means that both $N$ and the source-integrated electron spectrum $\langle nVF \rangle (E)$ are now finite, corresponding to a finite value of $E_{\rm min}$ in Equation~(\ref{eq: fbar-solution}).

To estimate the value of $E_{\rm min}$, we proceed as follows. At thermal energies $E \sim kT$ the approximate solution of Equation~(\ref{eq: fbar-solution}) is

\begin{equation}\label{eq: fbar-solution_approx}
\langle nVF \rangle (E) \simeq \frac{3\sqrt{\pi}}{2K} \, \sqrt{\frac{kT}{E_{\rm min}}} \, E \, e^{-E/kT} \, \int_{E}^{\infty} A \, F_0(E_0) \, dE_0
\simeq \frac{3\sqrt{\pi}}{2K} \, \sqrt{\frac{kT}{E_{\rm min}}} \, E \, e^{-E/kT} \, \dot{N} \, ,
\end{equation}
which has a Maxwellian form. Comparing this to the normalized expression for a Maxwellian with number density $n$ (cm$^{-3}$) and temperature $T$, viz.

\begin{equation}\label{eq: fbar-max}
\langle nVF \rangle (E) = \sqrt{\frac{8}{\pi m_e}} \, \frac{n N}{(kT)^{3/2}} \,  E \, e^{-E/kT} \,\,\, ,
\end{equation}
we see that $N$, the total number of thermalized electrons in the target, and ${\dot N}$, the injection rate of non-thermal electrons into the target, are related by

\[
\frac{3\sqrt{\pi}}{2K} \, \sqrt{\frac{kT}{E_{min}}} \, \dot{N}=\sqrt{\frac{8}{\pi m_e}} \, \frac{nN}{(kT)^{3/2}}
\]
or
\begin{equation}\label{eq:N-solution}
{\dot N} = \frac{2}{\sqrt{\pi}} \, \sqrt{\frac{E_{\rm min}}{kT}} \, \frac{N}{\tau_e} \,\,\, .
\end{equation}
Finally, comparing the diffusion result (Equation~(\ref{esc-diffusion})) with Equation~(\ref{eq:N-solution}),
we obtain an approximate expression for $E_{\rm min}$:

\begin{equation}\label{eq:E_min_basic}
\frac{E_{\rm min}}{kT}= \frac{\pi \, (kT)^2 \, \tau_e^4}{L^2 \, m_e^2} \simeq 625 \left ( \frac{kT}{\sqrt{2Kn L}} \right )^8 \,\,\, .
\end{equation}
This can be written in the equivalent form

\begin{equation}\label{eq:E_min}
\frac{E_{\rm min}}{kT}\simeq \left ( \frac{5 \lambda}{L} \right )^4 \,\,\, ,
\end{equation}
where $\lambda = (kT)^2/2Kn$, the distance required to stop an electron of energy $kT$ in a cold target. This quantity is proportional to the collisional mean free path, and we shall henceforth refer to it as the ``mean free path.''

\begin{figure}
\centering
\includegraphics[width=0.65\linewidth]{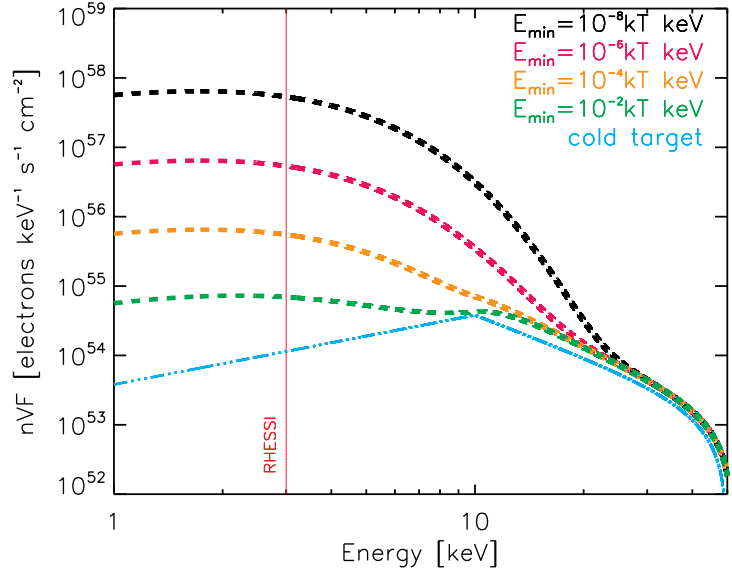}
\caption{The spatially integrated electron spectra $\langle nVF \rangle (E)$ given by the solution (\ref{eq: fbar-solution}) for various values of $E_{\rm min}$ given by Equation~(\ref{eq:E_min}), for a temperature and number density of $T=20$~MK and $n=10^{11}$~cm$^{-3}$, respectively. The injected spectrum $AF_0(E_0)$ is a power law with $\delta = 4$ and an injection rate $\dot{N} = 10^{36}$~s$^{-1}$ above $10$~keV.
The cold thick-target model is represented by the light blue line, and the vertical line at 3~keV indicates the lower limit of the {\em RHESSI} observation range.}
\label{fig:emin_test}
\end{figure}

The value of $E_{\rm min}$ is very sensitive to the target temperature $T$ ($E_{\rm min} \sim T^9$). It also depends inversely as the fourth power of the column density $\xi \equiv n L$ in the high temperature (coronal) region of the target.  This shows that, unlike in the cold-target model (in which the relationship between $F_0(E_0)$ and $\langle n V F \rangle (E)$ is independent of the spatial structure of the source - see Equation~(\ref{fbar-f0-cold})), the extent $L$ of the warm target region directly influences the relationship between the injected electron spectrum $F_0(E_0)$ and the source-integrated electron spectrum $\langle n V F \rangle (E)$.  In particular, for sufficiently extended coronal sources, $E_{\rm min}$ is very small and so from Equation~(\ref{eq: fbar-solution}), $\langle nVF \rangle (E)$ is greatly enhanced compared to its cold-target value (see Figure \ref{fig:emin_test}).

The effect of thermalization is particularly important when the hot plasma region can stop a large fraction of the injected electrons, i.e., when $E_{\rm c}^2 \lapprox 2KnL$ ($E_{\rm c}$ being the lower cutoff energy in the injected electron spectrum; see Equation~(\ref{f0pwrlaw}) below). In the opposite limit, $E_{\rm c}^2 \gapprox 2KnL$, injected electrons will reach the cold chromosphere without significant thermalization. For a large injection rate $\dot{N}$, the thermalization of injected electrons could be so significant such as to directly increase the density of the target. Specifically, if the deduced value of $\dot{N}$ satisfies

\[
\frac{\tau_e \, \dot{N}}{nAL} \gtrsim \frac{2}{\sqrt{\pi}} \, \sqrt{\frac{E_{min}}{kT}} \sim \frac{2}{\sqrt{\pi}} \, \left ( \frac{5 \lambda}{L} \right )^2 \,\,\, ,
\]
the plasma heating by the injected electrons needs to be taken into account. In other words, the minimum value of the low-energy limit $E_{\rm min}$ is

\begin{equation}
\frac{E_{\rm min}}{kT}\gtrsim \frac{\pi}{4} \left(\frac{\tau_e \dot{N}}{nAL}\right)^2 \,\,\, ,
\end{equation}
where $AL$ is the volume of hot plasma.

\subsection{Numerical Simulations}\label{sim_method}

In order to ascertain the accuracy of the analytic results above, we performed simulations using the methodology of \citet{2014ApJ...787...86J}, which includes the effects of both energy and pitch angle diffusion (see Equation~(\ref{eq:fp})).  Mimicking a real solar flare, we considered contiguous regions containing warm (coronal and/or flare-heated chromospheric) and cold (chromospheric) plasmas, respectively
(see left panel in Figure \ref{sims_bound}):

\begin{enumerate}
\item{Cold plasma at $|z|> L$, the temperature $T$ falls to a low $T<1$ MK and the density rises to $1\times10^{12}$ cm$^{-3}$};
\item{Hot plasma at $|z| \le L$, where warm target effects are important. }
\end{enumerate}

Beyond $|z| = L$ the background represents a region with properties similar to that of the lower chromosphere during a flare, with a temperature $T$ much lower than that of the flaring corona and a much higher number density $n$, that can collisionally stop electrons in the energy range in question over a very short distance $\ll 1 \arcsec$. In this region, electrons with energies less than 1~keV are removed from the simulation, since at such low $T$ and high $n$, it is unlikely that they will find themselves back in the coronal region between $-L<z<L$.

The resulting stationary solution is then given by the mean electron flux spectrum ${\langle n V F(E)\rangle}=A \sum nF(E) \, \Delta z$, where $\Delta z$ is the bin size and A is the cross-sectional area of the loop. The sum is over all simulation steps, for all $z$ and $\mu$ (including electrons in the chromospheric
region $|z|>L$ with $E>1$~keV).

\subsubsection{Simulation results}\label{sim-results}

For all simulation runs, we injected $10^4$~electrons, with an injected electron spectrum of the form

\begin{equation}\label{f0pwrlaw}
F_0(E_0) = \begin{cases}
0 &; E_0 < E_{\rm c} \\
C E_0^{-\delta} &; E_{\rm c} \le E_0 \le E_{\rm max}\\
0 &; E_0 > E_{\rm max}
\end{cases}
\end{equation}
with spectral index $\delta=4$, and low and high energy cutoffs of 10 and 50~keV, respectively. As a reference calculation, we used a coronal background temperature of $T=20$~MK and a background number density $n=1\times10^{11}$ cm$^{-3}$ within a region $|z| \le L = 20\arcsec$, so that the actual length of the region within the bounds is $40\arcsec$. This is large enough to contain the thick target coronal X-ray sources of $30\arcsec$ or so as seen in {\em RHESSI\,} observations \citep[e.g.,][]{2008ApJ...673..576X}. The temperature distribution $T(z)$ and number density distibution $n(z)$ over the entire region is shown in the left panel of Figure~\ref{sims_bound}.  For details of the simulation method see \citet{2014ApJ...787...86J}.

\begin{landscape}
\begin{figure*}[pht]
\centering
\includegraphics[width=0.32\linewidth]{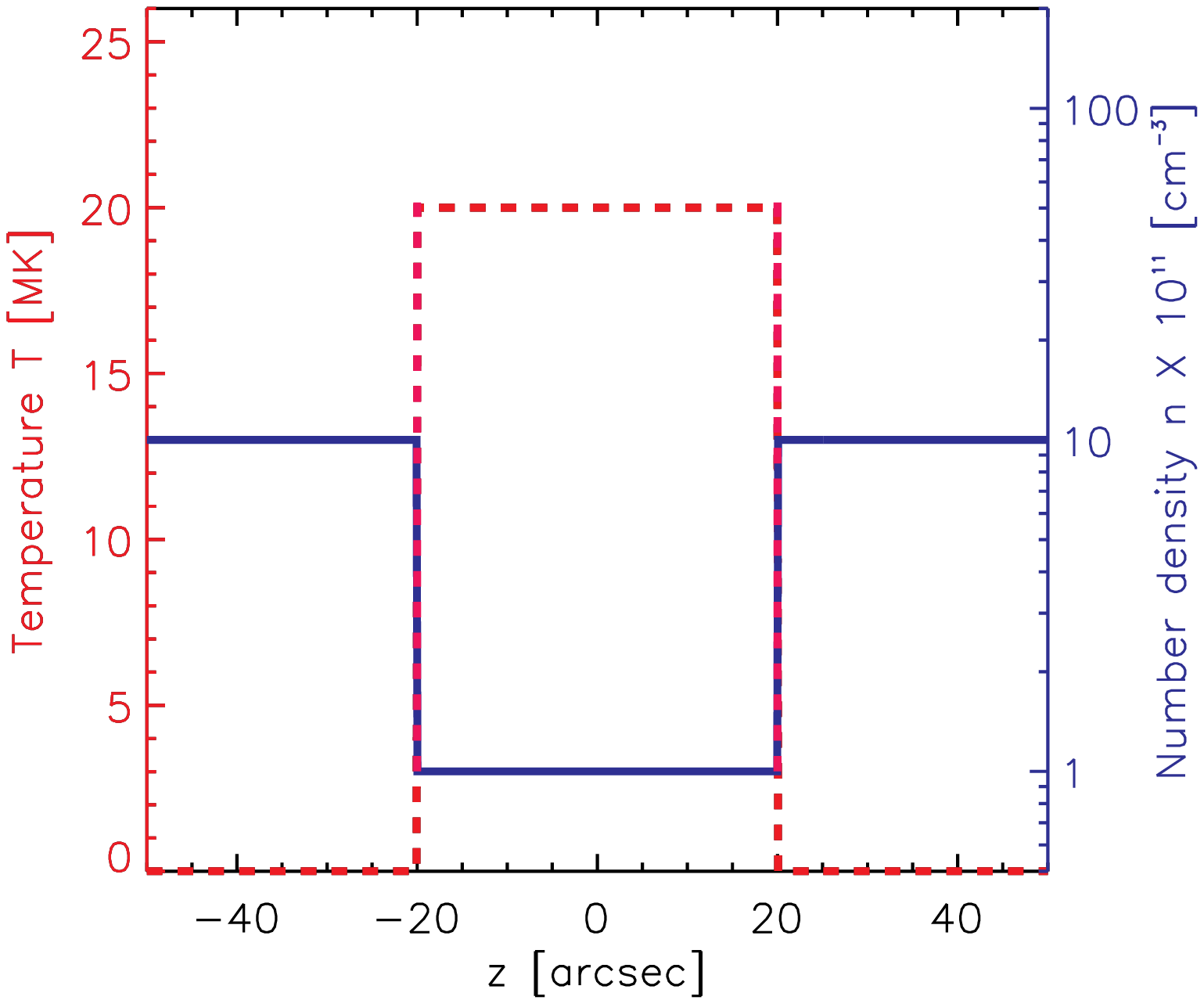}
\includegraphics[width=0.32\linewidth]{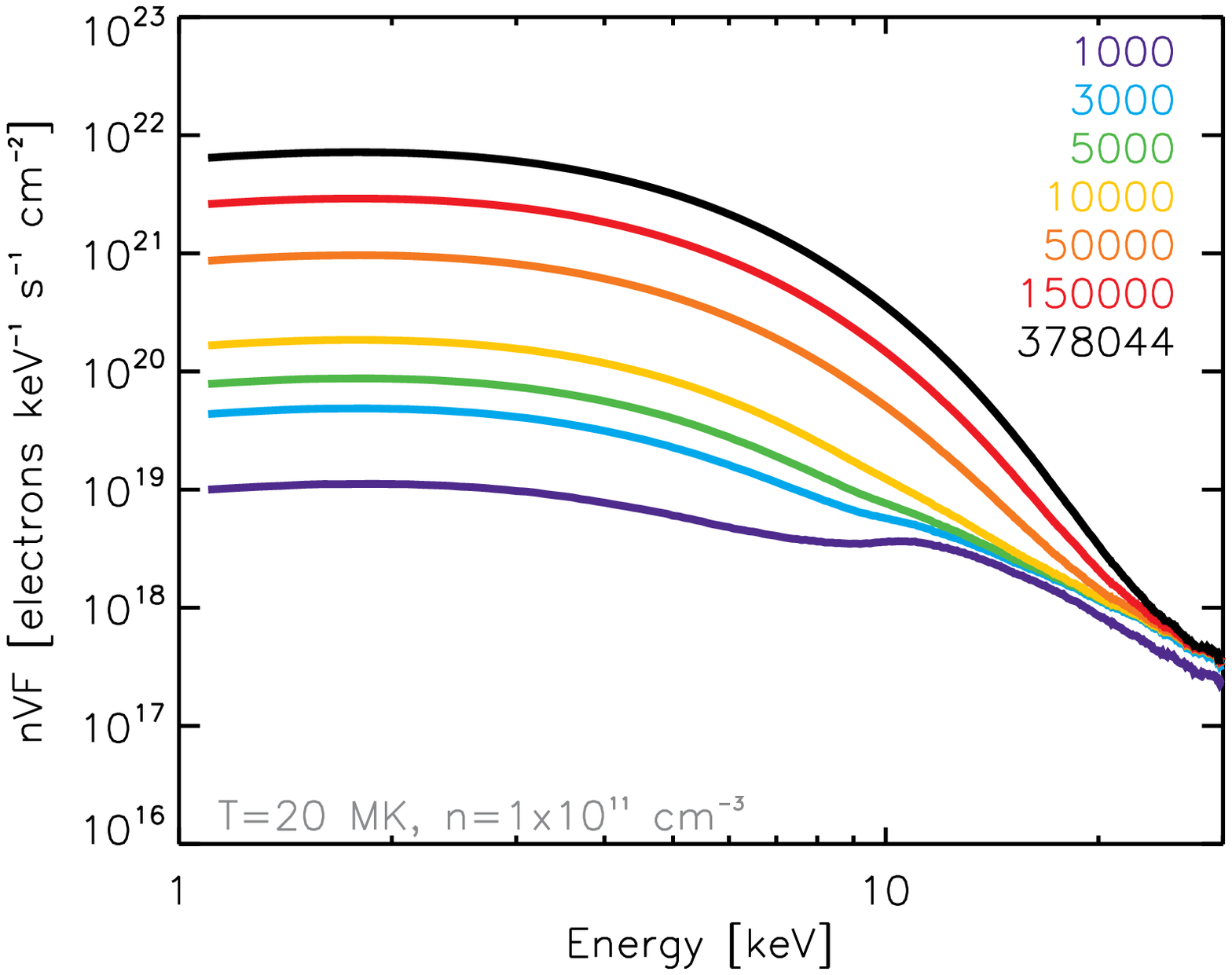}
\includegraphics[width=0.32\linewidth]{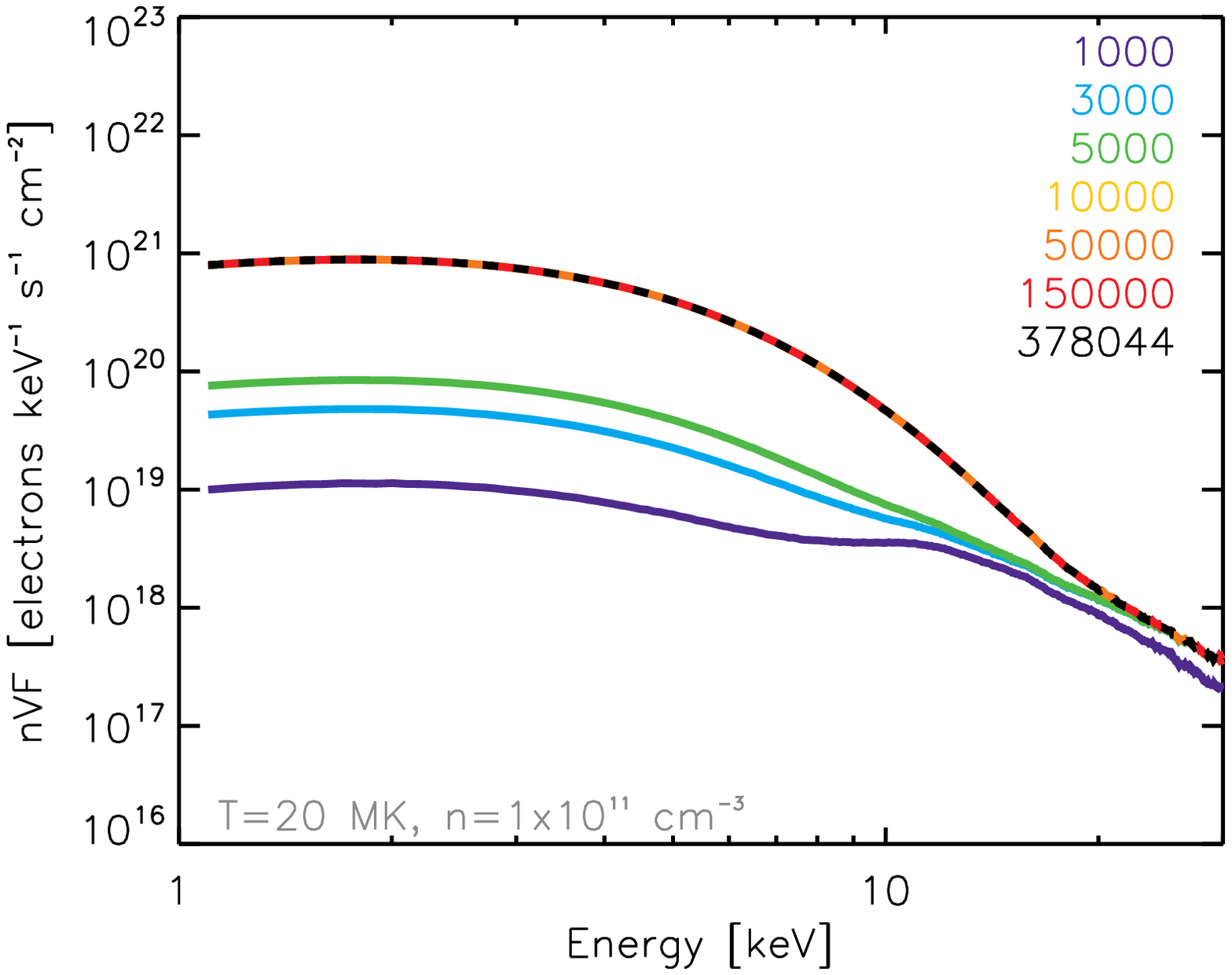}
\caption{The source-integrated electron spectra $\langle nVF \rangle (E)$ for a simulation with $T = 20$~MK, $n = 1 \times 10^{11}$~cm$^{-3}$, $\delta = 4$, an injected electron source size of $1\arcsec$ and an injection rate $\dot{N} = 1$~s$^{-1}$. \textit{Left}: The model temperature $T(z)$ and number density $n(z)$ distributions. \textit{ Middle}: No boundary conditions are imposed and hence the total integrated number of electrons continuously grows as the number of simulation steps increases. Hence, a stationary solution cannot be obtained. \textit{Right}: Using the same simulation, but now applying the boundary conditions listed in Section \ref{sim_method}; a stationary solution is now produced. The numbers in the middle and right panel legends denote the spatially integrated spectrum at the given simulation ``step.''}
\label{sims_bound}
\end{figure*}
\clearpage
\end{landscape}

\begin{figure*}[pht]
\centering
\includegraphics[width=\linewidth]{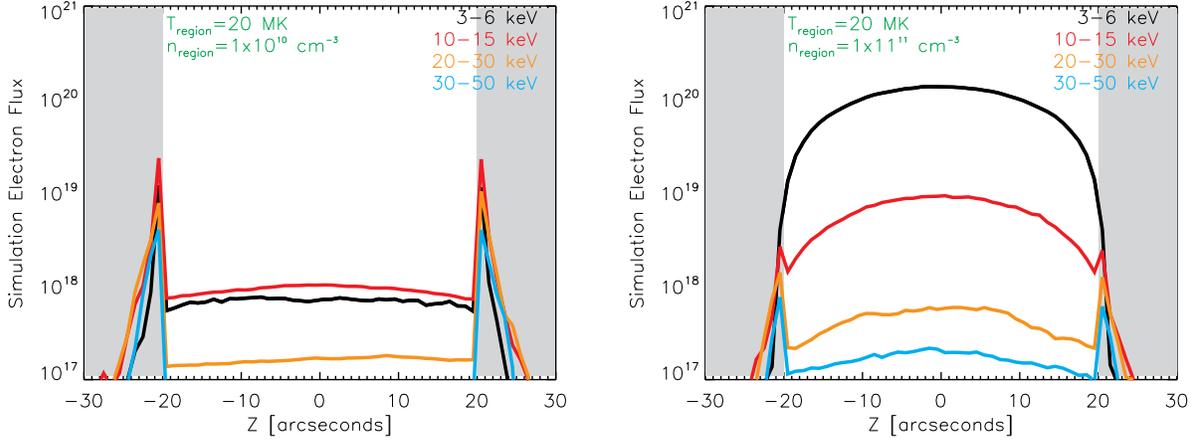}
\caption{The spatial distribution of X-ray emission along the coronal and chromospheric parts of the loop for two plasma densities of $10^{10}$~cm$^{-3}$ and $10^{11}$~cm$^{-3}$. The grey area delineates the chromospheric part of the loop. The parameters of the injected electrons are the same as in Figure~\ref{sims_bound}. }
\label{sims_fig_x}
\end{figure*}

The middle panel of Figure~\ref{sims_bound} shows a simulation with no boundary conditions imposed; the ever-increasing growth of the number of particles in the target is evident. By contrast, the right panel of Figure~\ref{sims_bound} shows the simulation results with the boundary conditions above imposed. The electron flux distribution reaches a stationary solution long before the end of the simulation. Figure~\ref{sims_fig_x} shows the resulting spatial distribution for two different simulation runs where the only difference is the coronal number density in the region $|z|<L$. Lower coronal densities lead to stronger footpoint emission.

\subsubsection{Effect of varying the low-energy cutoff}\label{vary-cutoff}

\begin{landscape}
\begin{figure*}
\centering
\includegraphics[width=0.49\linewidth]{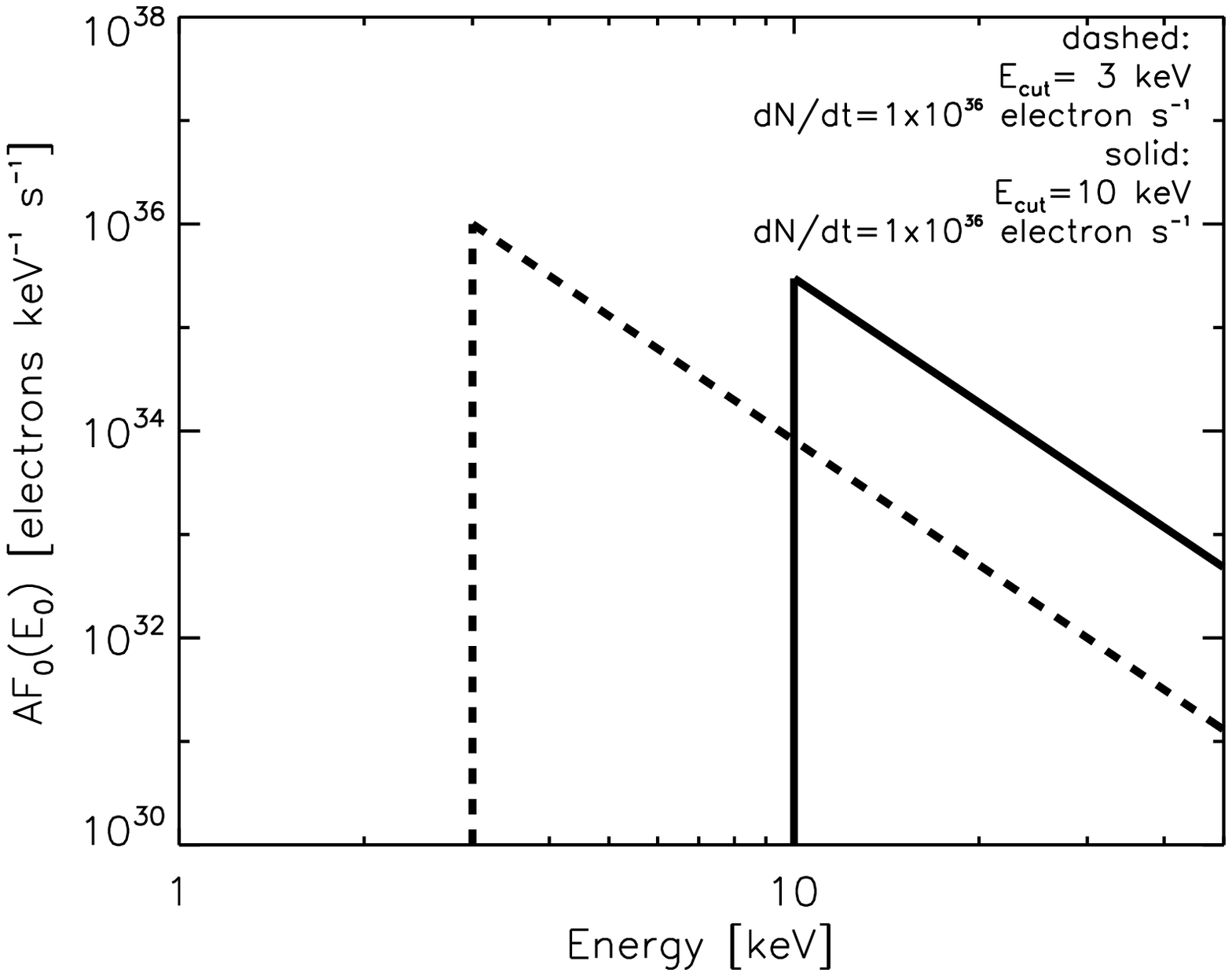}
\includegraphics[width=0.49\linewidth]{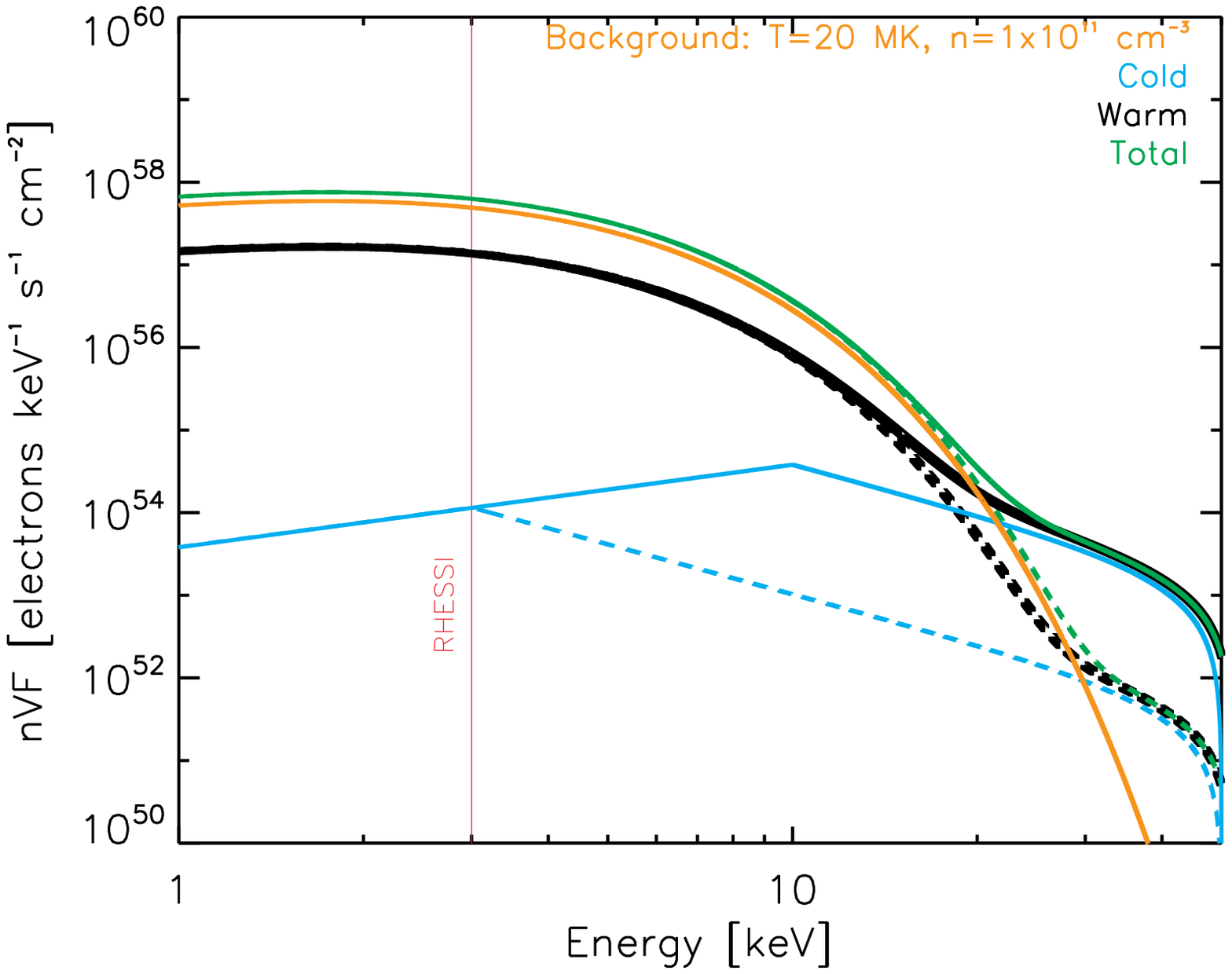}
\caption{\textit{Left panel}: Two different injected electron source distributions $F_0(E_0)$: (1) a power law with $\delta=4$ and $E_{\rm c}=3$~keV (dashed), and (2) a power law with $\delta=4$ and $E_{\rm c}=10$~keV (solid). Both have the same total number of injected electrons $\dot{N} = 1 \times 10^{36}$ electrons~s$^{-1}$. \textit{Right panel}: the resulting source-integrated electron flux spectra $\langle nVF \rangle (E)$, obtained from Equations~(\ref{eq: fbar-solution}) and~(\ref{eq:E_min}), for the case $T=20$~MK, $n=1\times10^{11}$~cm$^{3}$, $L=20\arcsec$, source volume $V = 10^{27}$~cm$^{3}$, corresponding to the background distribution shown in orange. Results for $\langle nVF \rangle (E)$ are denoted by the black dashed curve (spectrum (1)) and the black solid curve (spectrum (2)). The cold target results are shown in blue (dashed and solid). The total distributions (black plus orange) are shown by the green (dashed and solid) curves. The lower limit of the {\em RHESSI} data (at $E \simeq 3$~keV) is shown as a vertical line.}
\label{diff_ecut}
\end{figure*}
\end{landscape}

In the left panel of Figure~\ref{diff_ecut} we show two different injected electron distributions $F_0(E_0)$: a truncated power law with $\delta = 4$, $E_{\rm c} = 3$~keV and a truncated power law with $\delta = 4$, $E_{\rm c} = 10$~keV. In  the right panel of Figure~\ref{diff_ecut}, the resulting mean electron flux spectra $\langle nVF \rangle (E)$ are calculated for both cases using Equations~(\ref{eq: fbar-solution}) and~(\ref{eq:E_min}). The cold target results are also plotted for comparison, as is the thermal background distribution $\langle nVF \rangle_{\rm th} (E)$ for a source volume of $V = 10^{27}$~cm$^{3}$. The right panel of Figure~\ref{diff_ecut} also shows the overall total $\langle nVF \rangle (E) = \langle nVF \rangle_{\rm beam}(E) + \langle nVF \rangle_{\rm th}(E)$. While the solutions are significantly different at energies $E \gapprox 10$~keV, they are remarkably similar at lower energies, a consequence of the thermalization process since the electron behavior is mostly controlled by background Maxwellian properties.

\subsubsection{Comparison with previous results}\label{previous-results-comparison}

\begin{figure*}[pht]
\centering
\includegraphics[width=0.65\linewidth]{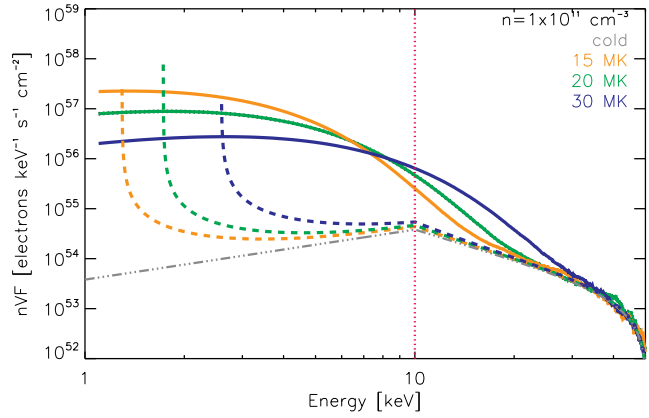}
\caption{Comparison of the spectra created from simulations with various physical characteristics (\textit{solid}: warm target including diffusion; \textit{dashed}: warm target, no diffusion; and \textit{grey dashed-dot}: cold target). All curves are normalized for an injection rate of ${\dot N} = 10^{36}$~s$^{-1}$. The plasma parameters are $T=15$~MK (orange), $T = 20$~MK (green) and $T = 30$~MK (blue), and $n = 1 \times 10^{11}$~cm$^{-3}$ in all cases. The initial electron distribution has a spectral index $\delta = 4$ and an injected energy range from $E_{\rm c} = $~10~keV to $E_{\rm max}=$~50~keV.  The vertical dashed line at 10~keV indicates the low energy cut-off.}
\label{sim_leff}
\end{figure*}

In Figure~\ref{sim_leff}, we compare the simulation results (using boundary conditions) with the results of \cite{2003ApJ...595L.119E}, where the case of a warm target without diffusion was studied analytically. The mean source electron flux spectrum is plotted for three target temperatures of $T = 15, 20$ and $30$~MK, using a number density of $n = 1 \times 10^{11}$~cm$^{-3}$; the cold target result is also included for comparison. The unrealistic case of a warm target without diffusion results in a large number of electrons accumulating at energies below that of the simulation thermal energy since such a model does not contain the physics to adequately describe the behavior of electrons $E\sim kT$ . Once diffusion is added, this feature disappears, since the energy can now diffuse about this point leading to the formation of a thermal distribution at lower energies. Hence the resulting spectrum has a large thermal component that dominates even above $E = 10$~keV, making the resulting spectral index steeper between 10 and 30~keV.

\begin{figure*}[pht]
\centering
\includegraphics[width=0.49\linewidth]{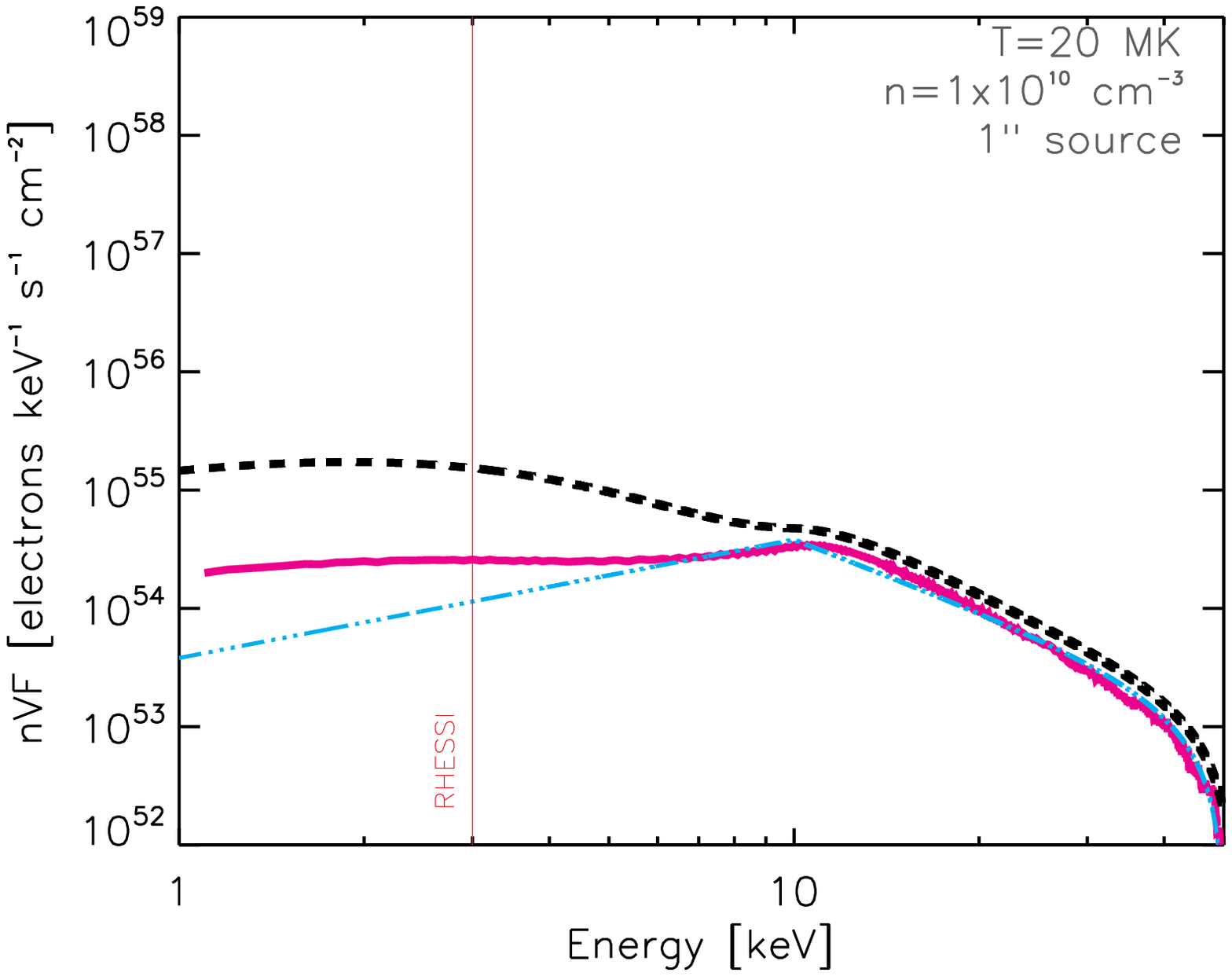}
\includegraphics[width=0.49\linewidth]{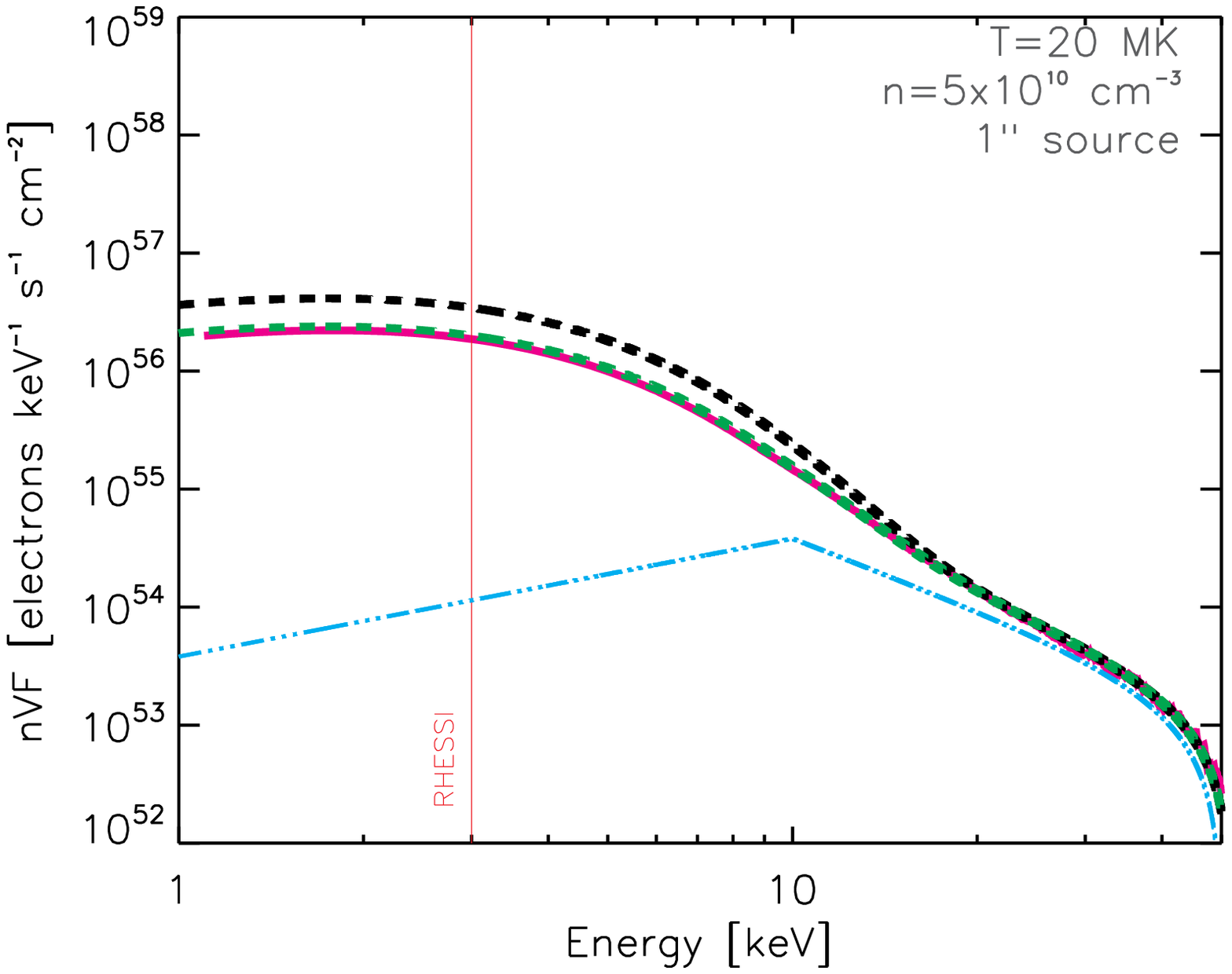}
\includegraphics[width=0.49\linewidth]{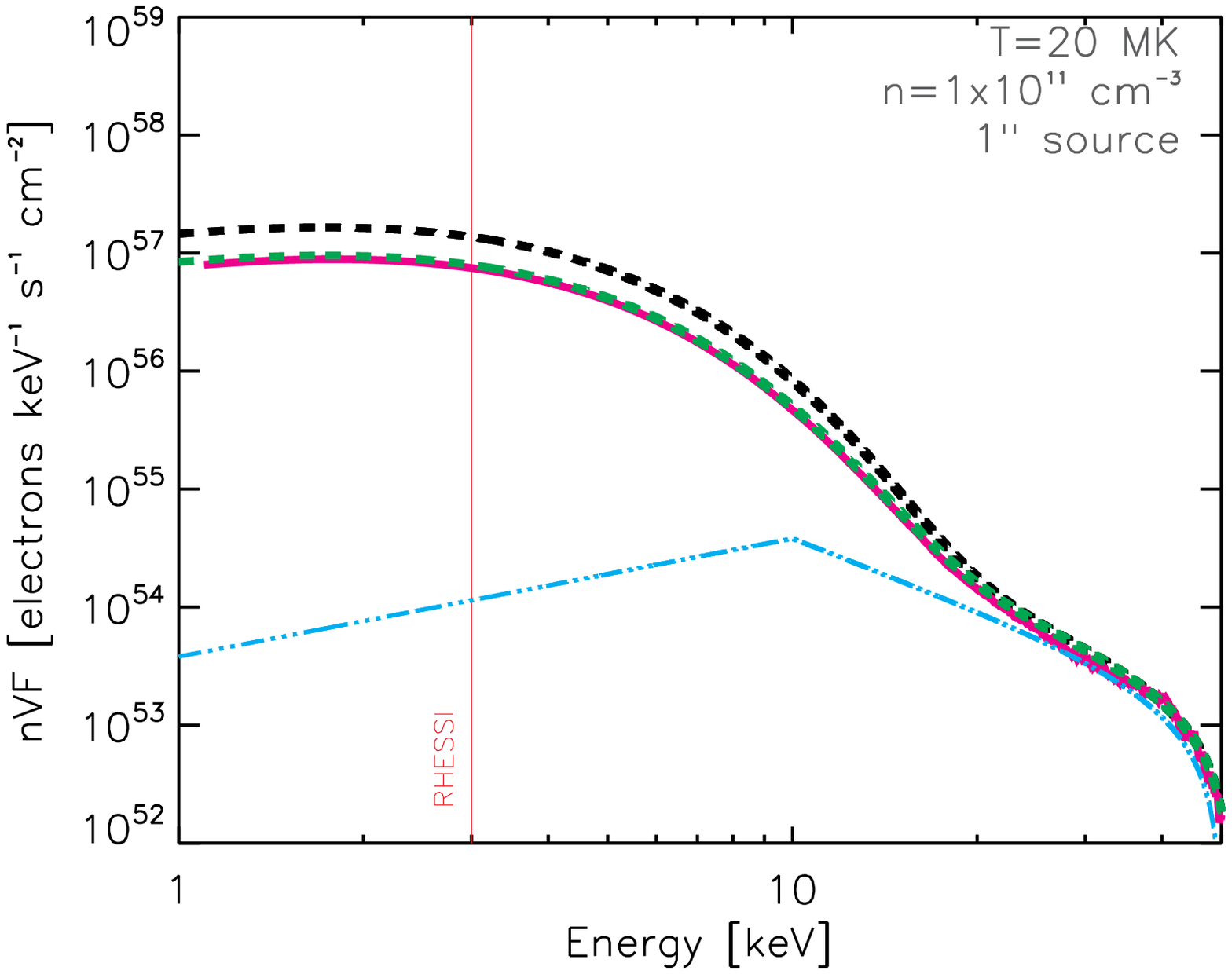}
\includegraphics[width=0.49\linewidth]{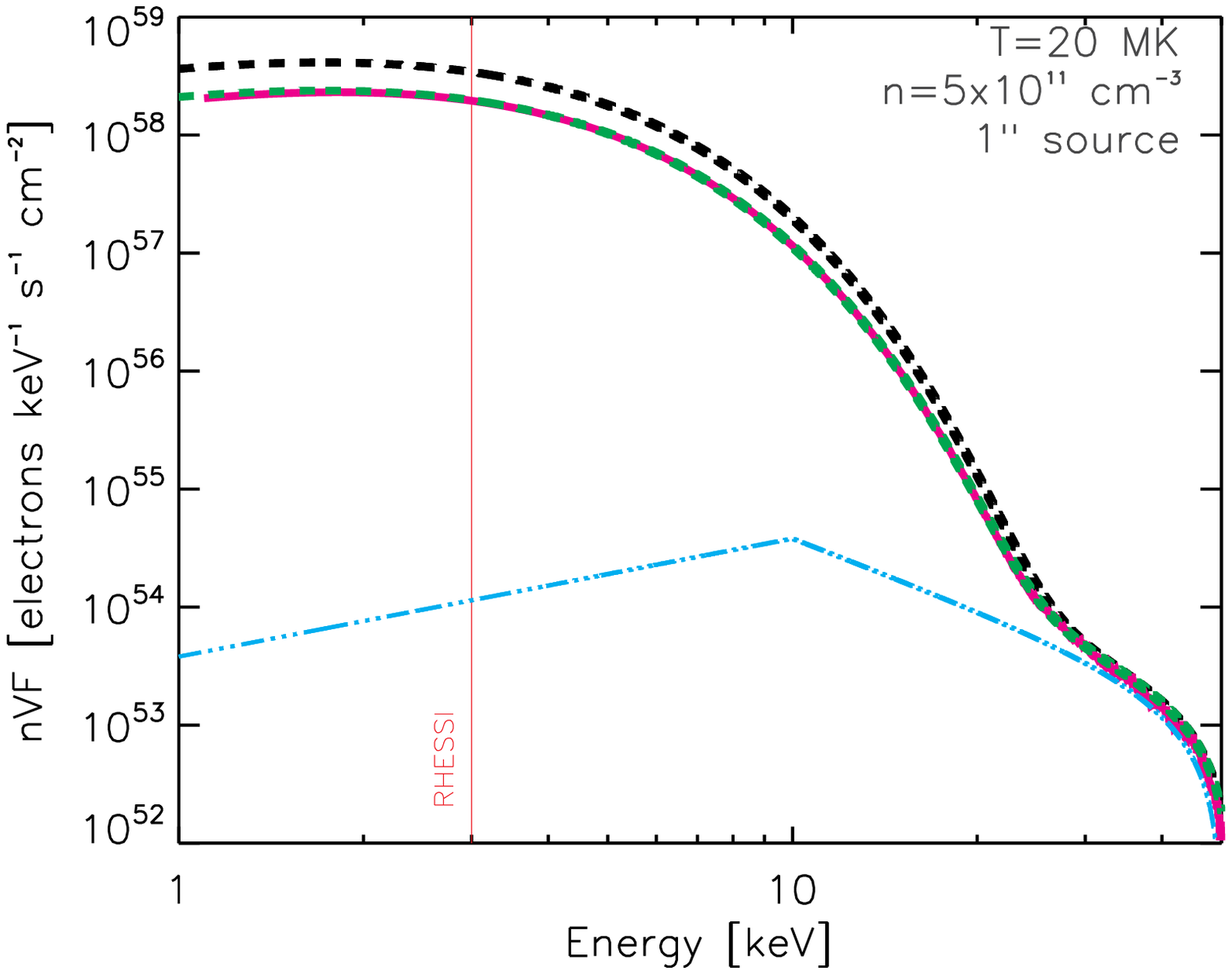}
\caption{Comparison of the analytic and simulation results for the mean source electron spectrum $\langle nVF \rangle (E)$. The background temperature is $T=20$ MK, with a (half) loop length of $20\arcsec$. The electron source parameters are $\delta=4$, $E_{\rm c} = 10$~keV, $E_{\rm max} =50$~keV and $\dot{N} = 10^{36}$ electrons~s$^{-1}$. Each plot has a different background density, from $n = 1 \times 10^{10}$~cm$^{-3}$ (\textit{top left}), $n = 5 \times 10^{10}$~cm$^{-3}$ (\textit{top right}), $n = 1 \times 10^{11}$~cm$^{-3}$ (\textit{bottom left}) and $n = 5 \times 10^{11}$~cm$^{-3}$ (\textit{bottom right}). \textit{Pink:} simulation, \textit{black:} analytical result (\ref{eq: fbar-solution}) with $E_{\rm min}$ given by Equation~(\ref{eq:E_min}), \textit{green (in applicable [high density] cases):} best match to the simulation using a value of $E_{\rm min}$ increased by a factor of 3 (see Equation~(\ref{eq:E_min_adj})), and \textit{light blue:} cold target.}
\label{sim_num_c1}
\end{figure*}

\begin{landscape}
\begin{figure*}[pht]
\centering
\includegraphics[width=0.32\linewidth]{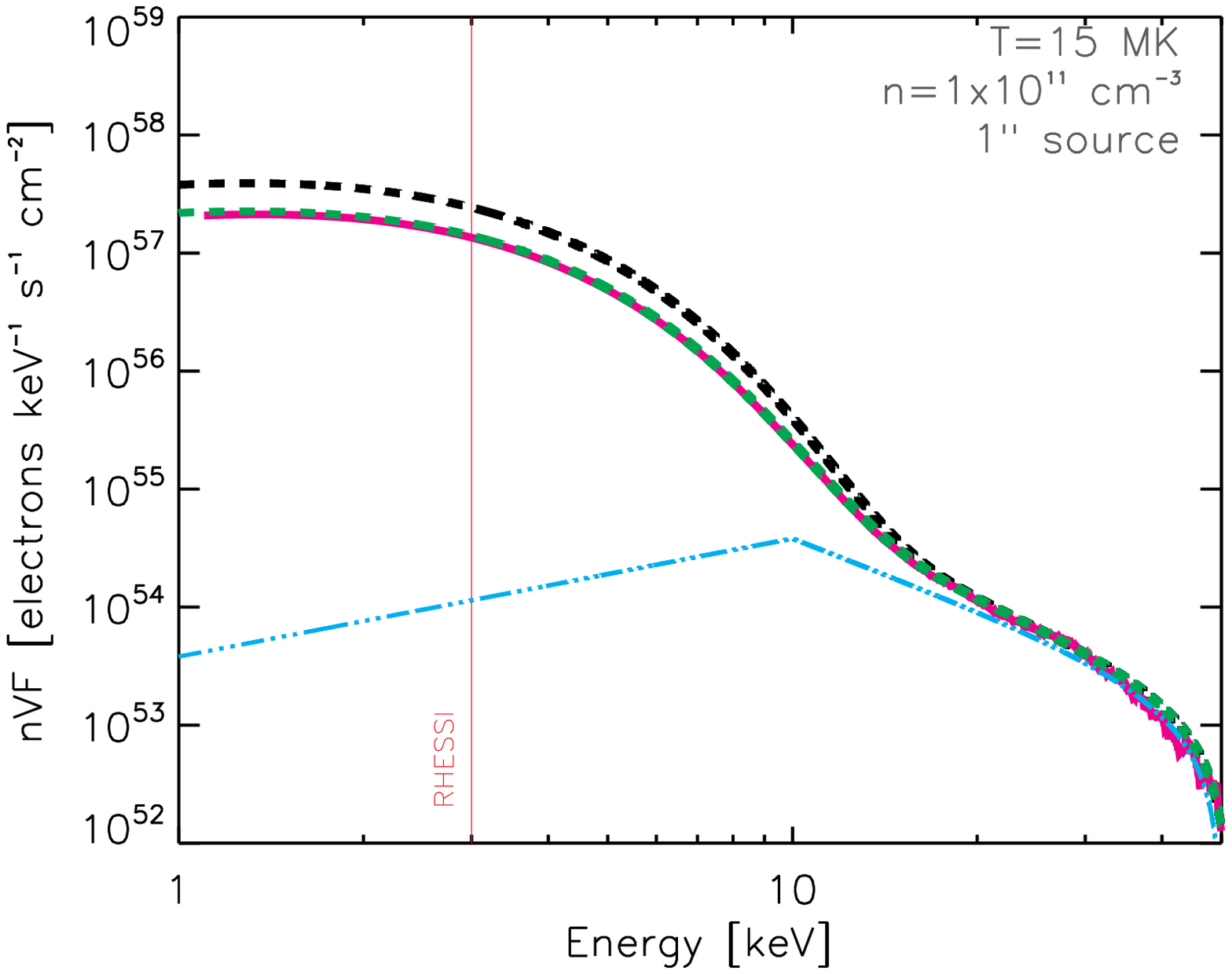}
\includegraphics[width=0.32\linewidth]{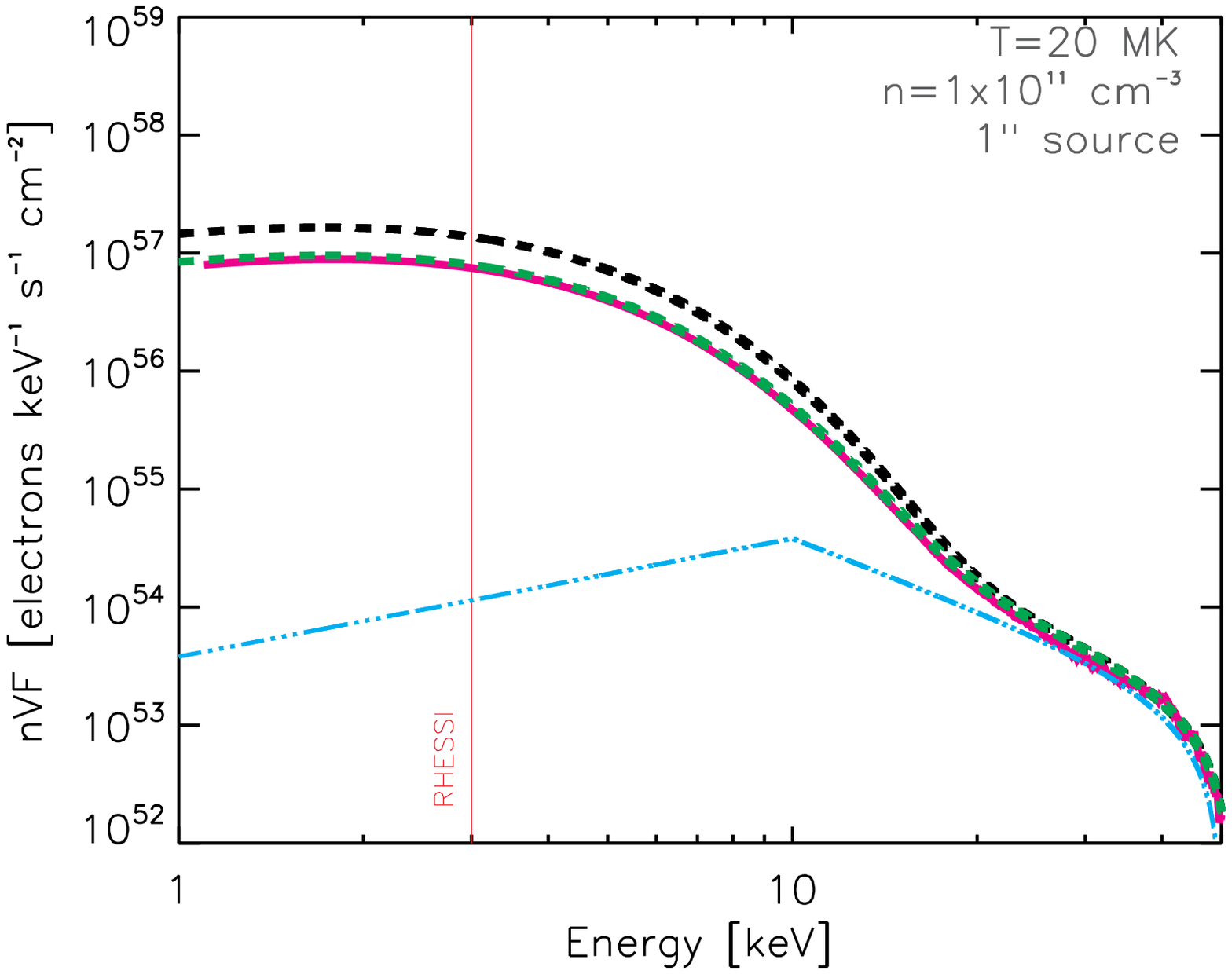}
\includegraphics[width=0.32\linewidth]{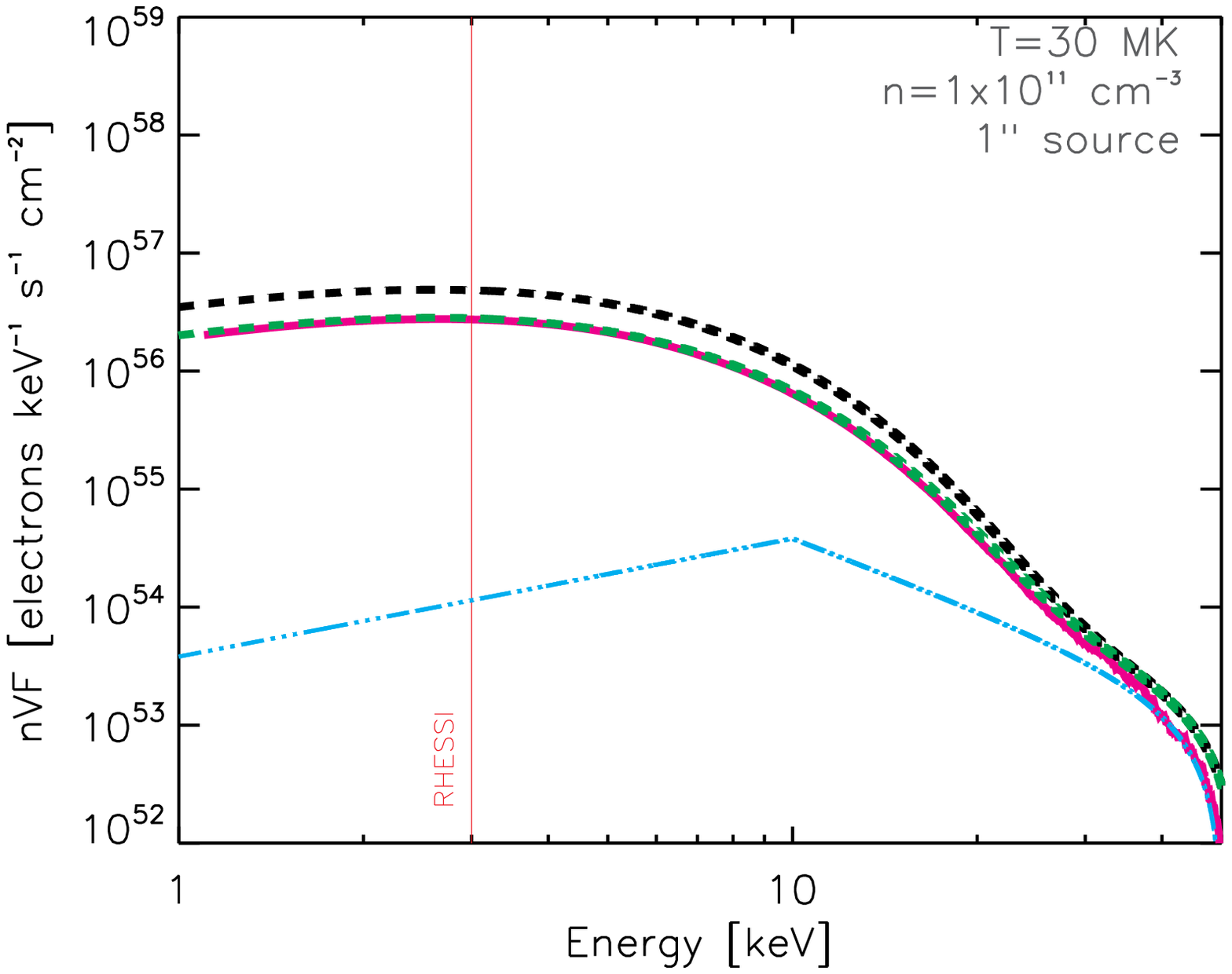}
\caption{As for Figure~\ref{sim_num_c1}, but a comparison between the simulated and analytical results is shown for different background temperatures: $T = 15$~MK (left), $T = 20$~MK (middle) and $T = 30$~MK (right).  In all cases the background density is $n = 10^{11}$~cm$^{-3}$. All other parameters are as noted in the Figure~\ref{sim_num_c1} caption.  \textit{Pink:} simulation, \textit{black:} analytical result (\ref{eq: fbar-solution}) with $E_{\rm min}$ given by Equation~(\ref{eq:E_min}), \textit{green:} best match to the simulation using a value of $E_{\rm min}$ increased by a factor of 3 (see Equation~(\ref{eq:E_min_adj})), and \textit{light blue:} cold target.}
\label{sim_num_c2}
\end{figure*}
\end{landscape}

\subsection{Comparison of analytical and numerical results}\label{comparison}

\begin{figure*}[pht]
\centering
\includegraphics[width=0.49\linewidth]{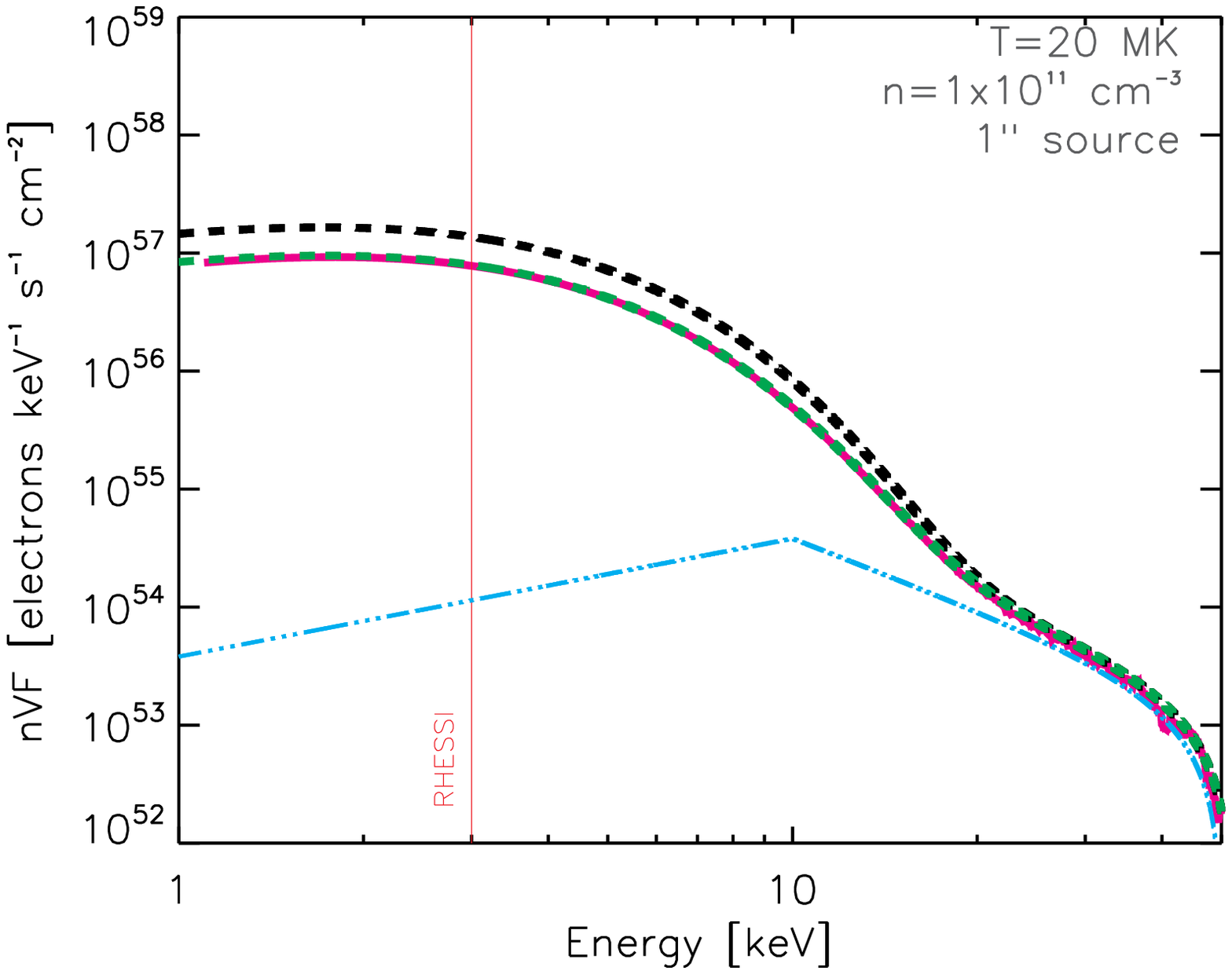}
\includegraphics[width=0.49\linewidth]{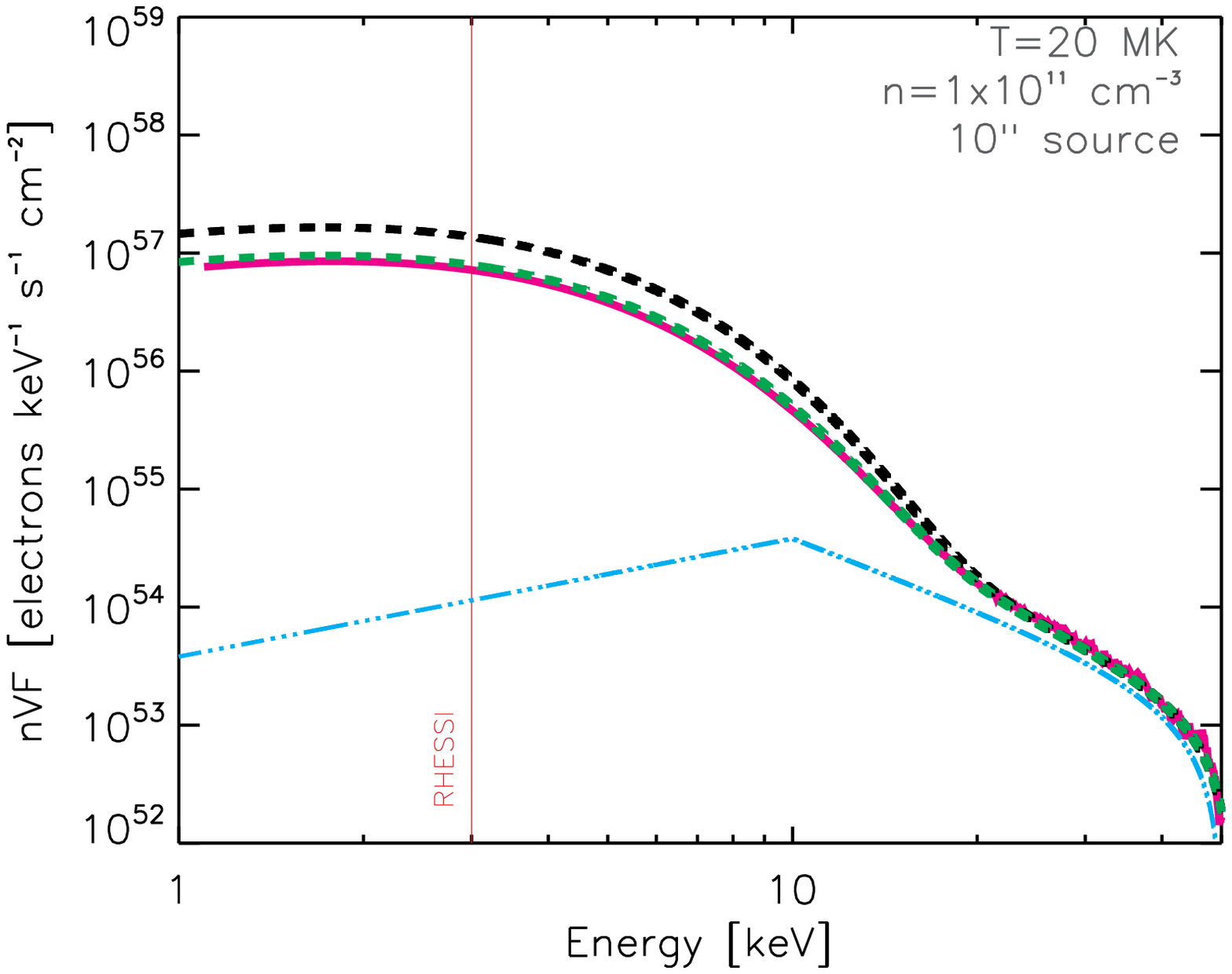}
\includegraphics[width=0.49\linewidth]{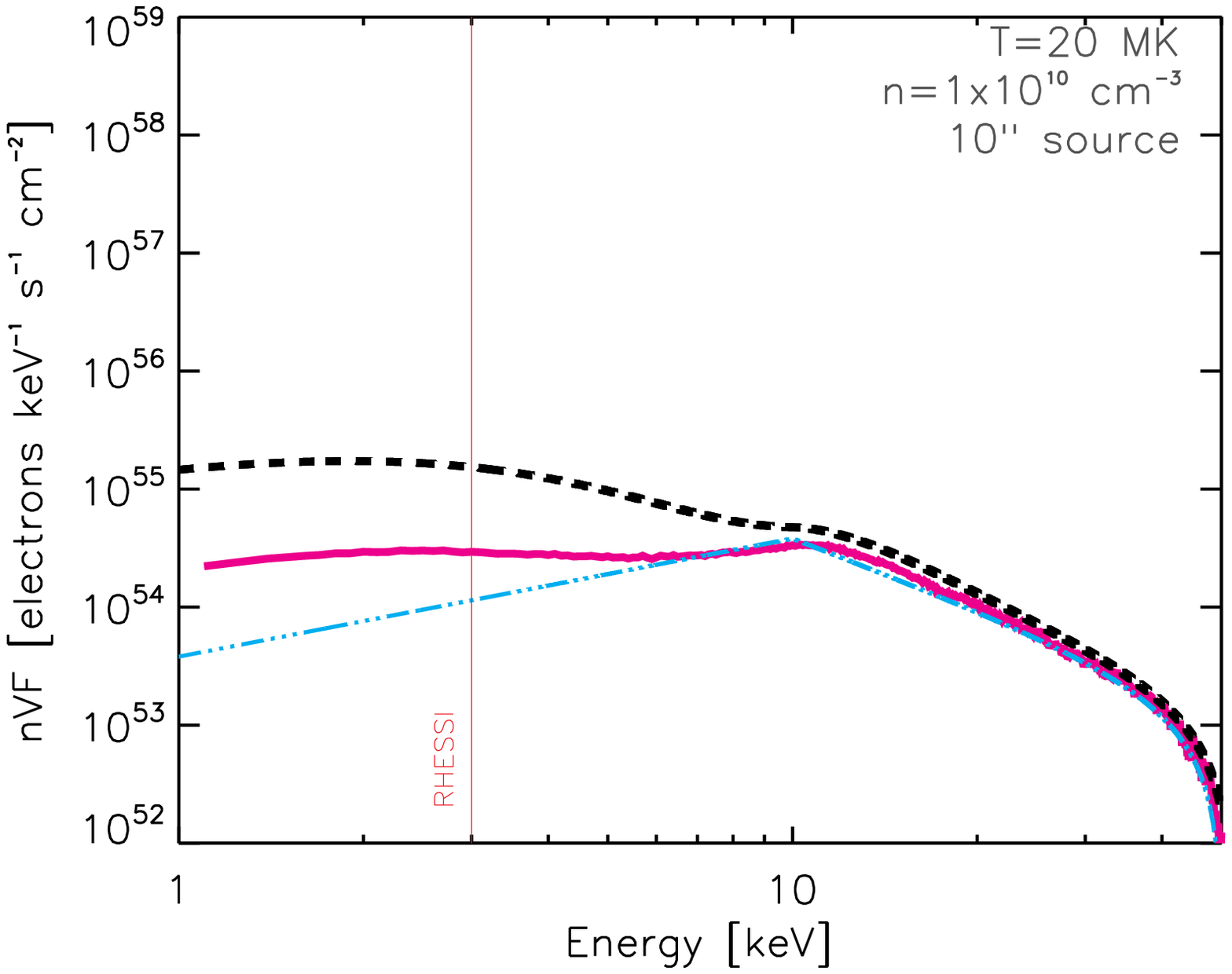}
\includegraphics[width=0.49\linewidth]{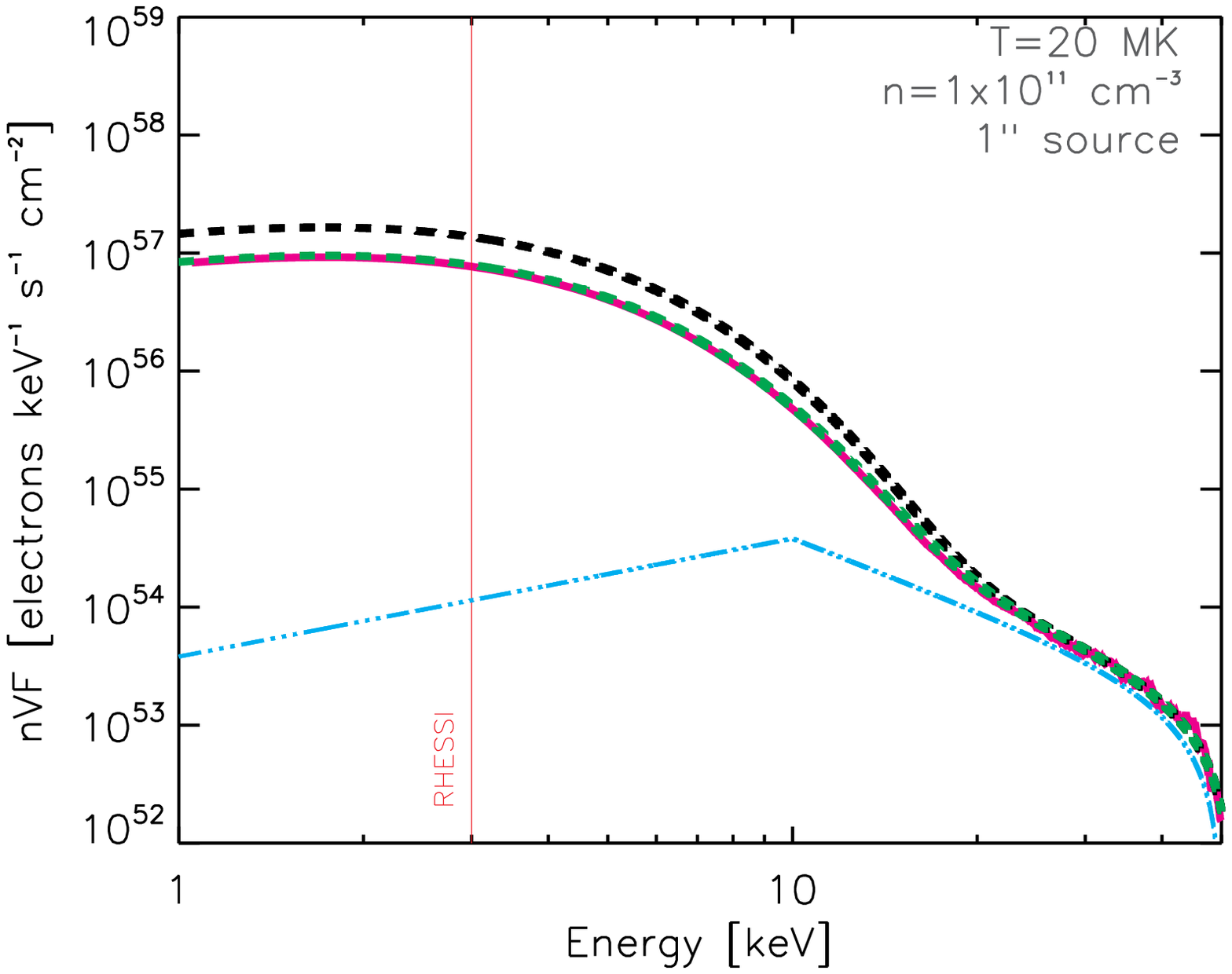}
\caption{As for Figures \ref{sim_num_c1} and \ref{sim_num_c2}, but for different electron source and boundary conditions: \textit{top left}: ($T=20$~MK, $n = 1 \times 10^{11}$~cm$^{-3}$) within the region, ($T = 8$~MK, $n = 1 \times 10^{12}$~cm$^{-3}$) outside the boundary, and an injected electron source size of $1\arcsec$, \textit{top right}: ($T=20$~MK, $n = 1 \times 10^{11}$~cm$^{-3}$) within the region, ($T = 8$~MK, $n = 1 \times 10^{12}$~cm$^{-3}$) outside the boundary and an injected electron source size of $10\arcsec$, \textit{bottom left} as for top right but with $n = 1 \times 10^{10}$~cm$^{-3}$ within the loop region, and \textit{bottom right}: ($T = 20$~MK, $n = 1 \times10^{11}$~cm$^{-3}$) in the region, ($T = 0.01$~MK, $n = 1 \times 10^{12}$~cm$^{-3}$) outside the boundary, an injected electron source size of $1\arcsec$, but using a pitch angle ($\mu$) distribution that is initially isotropic instead of initially beamed. \textit{Pink:} simulation, \textit{black:} analytical result (\ref{eq: fbar-solution}) with $E_{\rm min}$ given by Equation~(\ref{eq:E_min}), \textit{green (in applicable [high density] cases):} best match to the simulation using a value of $E_{\rm min}$ increased by a factor of 3 (see Equation~(\ref{eq:E_min_adj})), and \textit{light blue:} cold target.}
\label{sim_num_c3}
\end{figure*}

We can now evaluate the accuracy of the analytic solution of Section~\ref{finite_sol} (Equations~(\ref{eq: fbar-solution}) and~(\ref{eq:E_min})) by comparing it with the simulation results of Section~\ref{sim-results}. For this purpose, three sets of simulation runs were performed. In Figure~\ref{sim_num_c1} we compare the analytical and numerical solutions for a case with temperature $T=20$~MK, and for densities $n= 10^{10}$~cm$^{-3}$, $5 \times 10^{10}$~cm$^{-3}$, $ 10^{11}$~cm$^{-3}$ and $5 \times 10^{11}$~cm$^{-3}$, respectively. In Figure~\ref{sim_num_c2} we repeat this comparison for a plasma number density of $n = 10^{11}$~cm$^{-3}$ and three different temperatures ($T = 15$~MK, $T = 20$~MK and $T = 30$~MK). In Figure~\ref{sim_num_c3} we compare the numerical results with simulation runs using different plasma parameters beyond the boundary at $|z|=L$ and different electron source parameters. In the loop region, we again have $T=20$~MK and $n = 10^{11}$~cm$^{-3}$, but

\begin{enumerate}
\item{we inject the electron source over a $10\arcsec$ region instead of $1\arcsec$, or}
\item{we change the temperature beyond the spatial boundary $L$ to $T=8$~MK, or}
\item{we inject an electron distribution that is initially completely isotropic, instead of initially beamed.}
\end{enumerate}

\begin{figure*}[pht]
\centering
\includegraphics[width=0.65\linewidth]{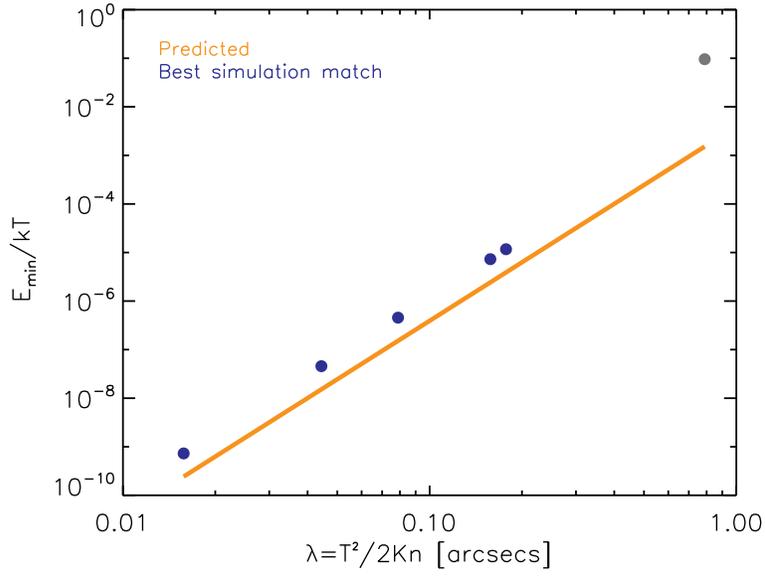}
\caption{The orange line shows the analytical-model value of $E_{\rm min}/kT$ (Equation~(\ref{eq:E_min})) versus the mean free path $\lambda$, while the blue dots show the $E_{\rm min}$ values that produce a mean electron flux spectra that best matches the results of the numerical simulations. It is apparent that setting $E_{\rm min}$ equal to approximately three times its analytic estimate (Equation~(\ref{eq:E_min})) produces a very good agreement between the analytic and numerical results.  For higher values of $\lambda$ (lower coronal densities), $E_{\rm min}$ begins to deviate much more from the analytic estimate (Equation~(\ref{eq:E_min}); grey dot), as the hot region moves into the thin-target domain.}
\label{emin}
\end{figure*}

Overall Figures~\ref{sim_num_c1}, \ref{sim_num_c2}, and~\ref{sim_num_c3} show that the analytic result matches the simulation results well, at least for all cases where the target density is high ($n \gapprox 3 \times 10^{10}$~cm$^{-3}$). However, as expected, the agreement becomes poorer for the low density cases (e.g., $n = 1 \times 10^{10}$~cm$^{-3}$) when the density of the loop region moves from a thick-target regime ($E_c^2 \ll 2KnL$) to a thin-target one ($E_c^2 \gg 2KnL$).  To explore this further, in Figure~\ref{emin} the predicted value of $E_{\rm min}/kT$ is plotted against the mean free path $\lambda$ (Equation~(\ref{eq:E_min})); the $E_{\rm min}$ values that best match the simulation results are also shown. The $E_{\rm min}$ values required to almost perfectly match the simulation results (for the high density ($n \gapprox 3 \times 10^{10}$~cm$^{-3}$) cases) are in general equal to $\sim 3$ times the value of $E_{\rm min}$ given by Equation~(\ref{eq:E_min}); this is an acceptable adjustment given the approximations made in Section \ref{finite_sol}.  (For low density cases, the value of $E_{\rm min}$ must be adjusted by a much larger factor; see grey dot in Figure~\ref{emin}.)  Analytic results using this revised value of $E_{\rm min}$ are shown, for the pertinent high-density cases, in Figures~\ref{sim_num_c1} through~\ref{sim_num_c3}.

In summary, the solution~(\ref{eq: fbar-solution}), repeated here:

\begin{equation}\label{eq: fbar-solution-repeat}
\langle nVF \rangle (E) = \frac{1}{2K} \, E \, e^{-E/kT} \, \int_{E_{\rm min}}^{E}
\frac{e^{E^\prime/kT} \, dE^\prime }{E^\prime \, G \left ( \sqrt{\frac{E^\prime}{kT}} \, \right ) } \,
\int_{E^\prime}^\infty A \, F_0(E_0) \, dE_0 \,\,\, ,
\end{equation}
with an adjusted value of $E_{\rm min}$ given by

\begin{equation}\label{eq:E_min_adj}
\frac{E_{\rm min}}{kT}\simeq 3 \, \left ( \frac{5 \lambda}{L} \right )^4 \,\,\, ,
\end{equation}
reproduces the results of a full Fokker-Planck simulation extremely well, at least for relatively high density ($n \gapprox 3 \times 10^{10}$~cm$^{-3}$) sources.  It incorporates (see Figures~\ref{sim_num_c1}, \ref{sim_num_c2}, and~\ref{sim_num_c3}) the power-law form of the spectrum at high energies, the Maxwellian form at low energies, and the transition between these two limiting regimes.

\section{SUMMARY}\label{discussion}

Our aim in this paper was to understand the impact of collisional energy diffusion, and hence of thermalization in a target of finite temperature, on the deduction of the properties of accelerated electrons in solar flares. The results show that energy diffusion dramatically changes the energy evolution of electrons with energies $E \lapprox 10 \, kT$.

Via numerical simulations of the electron dynamics that take into account the effects of deterministic energy loss and diffusion in both energy and pitch angle, we have determined the form of the source-integrated electron spectrum $\langle nVF \rangle (E)$ for a variety of assumed injected spectra $F_0(E_0)$ and target models. We have also derived an analytic expression (Equations~(\ref{eq: fbar-solution-repeat}) and~(\ref{eq:E_min_adj})) for $\langle nVF \rangle (E)$ that is valid in the limit where most of the injected electrons are collisonally stopped in the warm target region, forming a thermally relaxed distribution which undergoes spatial diffusion while escaping to the cold chromospheric region. When $E_{\rm min}$ becomes comparable to $kT$, our model approximates the standard thick-target results. Overall, the predictions of this model compare favorably with the numerical results.

As shown in Section~\ref{analysis-analytical}, the use of the more accurate relation~(\ref{eq:nvF_full}) between the injected electron flux spectrum $F_0(E_0)$ and the observationally-inferred mean source electron spectrum $\langle nVF \rangle (E)$ results in an order of magnitude reduction of the deduced number of injected electrons at energies $E_0 \lapprox 10 kT$ compared to the cold target result~(\ref{fbar-f0-cold}) and even compared to the non-diffusional warm target result~(\ref{f0-fbar}). Use of a physically complete warm-target model leads to a lower bound on the value of the low-energy cut-off $E_{\rm c}$ and hence an upper bound on the total injected power ${\cal P} = A \, \int E_0 \, F_0(E_0) \, dE_0$. This provides a new alternative to the current practice \citep[e.g.,][]{2003ApJ...595L..97H} of identifying the {\it maximum} value of $E_{\rm c}$ that is consistent with the observed hard X-ray spectrum (in a cold-target approximation), and hence the determination of an {\it lower} bound on the total injected power ${\cal P} = A \, \int_{E_{\rm c}}^\infty E_0 \, F_0(E_0) \, dE_0$.

We therefore discourage the use of cold thick-target model, especially in cases of warm and relatively dense coronal sources. Instead, we advocate the use of the more physically complete target model, including the effect of electron thermalization. To derive the source-integrated electron spectrum $\langle nVF \rangle (E)$ (and so, from Equation~(\ref{xray-electron}), the hard X-ray spectrum $I(\epsilon)$) for a prescribed injected flux spectrum $F_0(E_0)$, use Equation~(\ref{eq: fbar-solution-repeat}) in conjunction with Equation~(\ref{eq:E_min_adj}).  The development of such fit model compatible with RHESSI software is currently underway.

\acknowledgments

The authors are grateful to Gordon Holman and Brian Dennis for helping to improve the manuscript. EPK and NHB gratefully acknowledge the financial support by an STFC Grant. NLSJ was funded by STFC and a SUPA scholarship. AGE was supported by NASA Grant NNX14AK56G.

\bibliographystyle{apj}
\bibliography{all_issi_references}

\appendix
\end{document}